\documentclass[twocolumn,showpacs,preprintnumbers,amsmath,amssymb]{revtex4}

\usepackage{graphicx}
\usepackage{dcolumn}
\usepackage{bm}
\usepackage{color,subfigure}

\begin{document}
\preprint{}

\title{Dynamic Screening in a Two-Species Asymmetric Exclusion Process}
\author{Kyung Hyuk Kim}
\affiliation{Department of Physics, University of Washington, Seattle,
WA 98195}
\author{Marcel den Nijs}
\affiliation{Department of Physics, University of Washington, Seattle,
WA 98195}
\date{\today}

\begin{abstract}
The dynamic scaling properties of the one dimensional Burgers
equation are expected to change
with the inclusion of additional conserved degrees of freedom. We
study this by means of
1-D driven lattice gas models  that conserve both mass and momentum.
The most elementary version of this is the Arndt-Heinzel-Rittenberg
(AHR) process, which is usually presented as
a two species diffusion process, with particles of opposite charge
hopping in opposite directions and with a
variable passing probability.
 From the hydrodynamics perspective this can be viewed as two coupled
Burgers equations, with the number of positive
and negative momentum quanta individually conserved.
We determine the dynamic scaling dimension of the AHR process from
the time evolution of the two-point correlation functions,
and find numerically that the dynamic critical exponent is
consistent with simple Kardar-Parisi-Zhang (KPZ) type scaling.
We establish that this is the result of perfect screening
of fluctuations in the stationary state.
The two-point correlations decay exponentially in our simulations and
in such a manner that in terms of quasi-particles,
fluctuations fully screen each other at coarse grained length scales.
We prove this screening rigorously using the analytic matrix product
structure of the stationary state.
The proof suggests the existence of a topological invariant.
The process remains in the KPZ universality class but only in the
sense of a factorization, as $(\mbox{KPZ})^2$.
The two Burgers equations decouple at large length scales due to the
perfect screening.
\end{abstract}
\maketitle

\section{Introduction}
Many non-equilibrium driven systems display scale invariance in their
stationary states, i.e.,
strongly correlated collective structures without a characteristic
length scale limiting the fluctuations.
Such correlations typically decay as power laws with critical
exponents that are universal.
Their values depend only on  global issues like dimensionality,
symmetry, and specific microscopic conservation laws.
The classification of dynamic universality classes and the
determination of their scaling dimensions is one
of the central issues in current research of non-equilibrium
statistical physics \cite{Hinrichsen00, Odor04}.
The one-species asymmetric exclusion processes (ASEP)  serves in this
context as both the simplest
prototype model for driven one-dimensional (1D) stochastic particle
flow \cite{Ligget85, Spohn91, Schmittmann95} and as a fully discretized
version of the 1D Burgers equation (with time and space discretized,
and momentum quantized) \cite{Halpin95}.

In this paper we investigate how the properties of such stochastic
flows change with the introduction
of additional  bulk conservation laws.
The generic expectation is that  enforcing more conservation laws
changes the scaling dimensions.
We follow a bottom-up  approach. An example of a top-down approach is the
current interest in anomalous 1D heat conduction in  Fermi-Pasta-Ulam
type models (e.g., a chain of anharmonic oscillators \cite{Lepri03}, or
a one dimensional gas of particles in a narrow channel with different
types of interactions \cite{Grassberger02}).
The systems are coupled to heat reservoirs on either end. Those  are
held at different temperatures and thus induce  heat flow along the channel.
Computer simulations, e.g., using  molecular dynamics, show an
anomalous thermal conductivity, $\kappa$,
$J_Q\simeq \kappa \nabla T$, diverging  with system size $L$ as
$\kappa \sim L^\alpha$.
The numerical estimates for the value of $\alpha$ in the various
versions of the process
vary between  $0.22<\alpha<0.44$ \cite{Lepri03, Grassberger02, Wang04}.
$\alpha$ is expected to be universal.
 From the analytic side, a  mode-coupling treatment predicted
$\alpha=2/5$ \cite{Wang04-2}, while
a renormalization analysis of the full hydrodynamic equations
predicts $\alpha=1/3$, based on Galilean invariance and an assumption
of local equilibrium
in the heat sector \cite{Narayan02}.
In our study we add conservation laws to the Burgers equation
instead of coarse graining down from full hydrodynamics.

The equivalences between ASEP, KPZ growth, and the Burgers equation
are well known  \cite{Halpin95}.
ASEP is usually interpreted as a process for stochastic particle
transport, while the Burgers equation
\begin{eqnarray}
\label{burgers}
\frac{\partial v}{\partial t} &=& \frac{\partial}{\partial x}
\left[\nu \frac{\partial v}{\partial x}+\lambda v^2+\eta(x,t)
\right]
\end{eqnarray}
represents the evolution of a (vortex free) velocity field $v(x,t)$,
and conserves
momentum only \cite{Burgers74}. The interpretation of ASEP as a fully 
discretized
Burgers equation poses some conceptional issues.
Due to the full quantization of the momenta in ASEP, in units of
$n=0,1$,  it can appear that the process
also conserves energy. A careful discussion \cite{Marcel} shows that 
energy is conserved
between updates but fluctuates during each update.
Therefore ASEP is a genuine
fully discretized implementation of the Burgers equation from this
direct point of view as well.
In section \ref{ahr} we discuss how to impose conservation of
particles in addition to conservation of momentum.
This leads naturally to the two-species ASEP known as the
Arndt-Heinzel-Rittenberg (AHR) model \cite{Arndt98, Arndt99}.
This process has been the focus of intensive studies, but
its dynamic scaling properties seem to have been ignored.
Instead, the stationary state properties have been center stage, in
particular  its clustering,
and that it can be constructed exactly using the  so-called matrix
product ansatz method \cite{Rajew00, Kafri02, Schutz03, Kafri03, Evans05}.

We establish that the introduction of this additional conservation
law to ASEP does not  change the universality class,
but it does so in a rather intricate manner. KPZ scaling changes to
(KPZ)$^2$ type scaling.
The AHR process can be interpreted as a coupled Burgers and
diffusion equation, conserving both mass
and momentum; or as  two coupled Burgers equations, one for positive
and negative momentum quanta separately.
The latter point of view turns out to be the most productive. At
large length scales the coupling
vanishes and the process factorizes, in terms of quasi-particles,
into two decoupled Burger processes.
This is achieved by means of perfect screening of fluctuations in the
stationary state. We observe this numerically from the
behavior of the two-point correlators (sections \ref{rispscreening}
and \ref{rnotpscreening}). The stationary state of the model is known
to satisfy
the so-called Matrix Product ansatz \cite{Arndt98}. We use that 
property to prove
analytically that the perfect screening is rigorous (section
\ref{MPA}).
In sections \ref{rispscreening} and \ref{rnotpscreening}
we present also direct numerical evidence that the dynamic critical
exponent is indeed the
same as in  KPZ, $z=3/2$, using the time evolution of the two  point
correlators. The conventional methods fail due to time oscillations.
This might be the first example of such a numerical dynamic analysis
in terms of correlation functions.

\begin{figure}[b]
\begin{center}
\includegraphics*[width=0.9\columnwidth, angle=0] {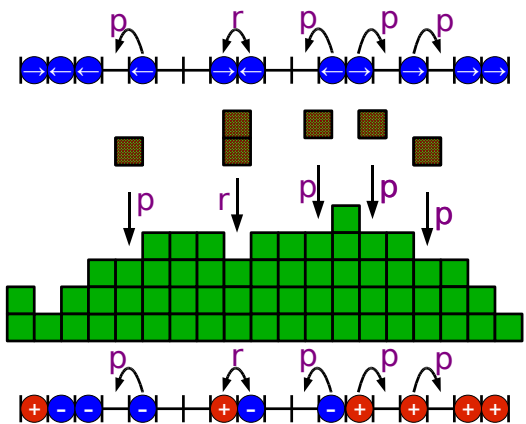}
\end{center}
\caption{\label{growth-asep} Two species asymmetric exclusion
process (bottom) and its corresponding interface growth model 
(middle) and particle flow model with momentum conservation (top).}
\end{figure}

\section{The AHR model}
\label{ahr}

Our aim is to construct a generalization of ASEP
describing a process where particle diffusion and the Burgers
equation are coupled to each other.
Energy will not be conserved.
The particles in such a model need to carry an internal degree of
freedom, representing momentum.
A site could be in four states. It would be empty ($n_x=0$) or be
occupied by a particle ($n_x=1$) with momentum $v_x=+1,0,-1$.
Particles with $+1$  (-1) momentum would hop with a right (left)
bias. Some reflection on the nature of the passing processes (the
collisions)
shows that we can  remove the zero momentum state of
particles, without loss of generality \cite{Marcel}.

This then leads naturally to the two-species ASEP known as the
Arndt-Heinzel-Rittenberg (AHR) model.
The conventional interpretation of this process is in terms of
diffusion  of charged particles in an electric field.
Two species of particles with opposite unit charge hop in opposite
directions along a 1D lattice ring,
driven by the electric field.
\begin{equation}
+0\overset{p}{\rightarrow} 0+  \qquad 0- \overset{p}{\rightarrow} -0
\qquad +- \overset{r}{\underset{t}{\rightleftarrows}} -+
\end{equation}
Each site $x$ can be in 3 states, $S_x=0,\pm 1$, with  $S=1$ ($S=-1$)
representing the right (left)
moving species and $S=0$ an empty site.
$p$ is the free directed hopping rate (the electric field) and $r$ the
passing rate of opposite charged particles.
In our study, the numbers of $S=1$ and $S=-1$ particles on the ring
are chosen to be equal.   Compared to the conventional single species ASEP,
this process has two local conservation laws instead of one;
both species are conserved independently.

In the coupled diffusion-Burgers equation interpretation of the same process,
the charge represents a quantum of momentum moving in opposite direction as
illustrated in Fig.\ref{growth-asep}. No driving force is present, because the
preferred hopping direction represents the total derivative in the
Navier-Stokes equation, just as in the single species ASEP.
Similar to ASEP, energy is not a conserved quantity:
The energy of particles  is conserved  between updates
but fluctuates during the updates. That leaves particles in
different places than where they would have been if energy were conserved
\cite{Marcel}.

The AHR model reduces to the $S_x=\pm 1$ spin (momentum quanta)
representation of ASEP in the high density limit where
vacant sites $S=0$ are absent. There, the particle density can not
fluctuate anymore, and the process falls thus back to the Burgers equation
with only one  conservation law.
This limit is singular. The AHR process is not
the generic $S=0,\pm 1$ generalization of ASEP in
the sense of the KPZ and Burgers equation. The proper generalization
would be the so-called restricted solid-on-solid (RSOS) model
(Kim-Kosterlitz model) where $+$ and $-$ pairs can be annihilated
and created. Those processes conserve momentum.
The $S=+1$ ($S=-1$) particles represent  up (down)  steps in the KPZ type
interface, the free hopping rate $p$ represents step-flow.
Growth at flat terraces is blocked in the AHR process, except for
the deposition of vertical dimers (with rate $r$)
in single particle puddles. Fig.\ref{growth-asep} illustrates this.

This means that from the  KPZ point of view the AHR process represents a
growing interface where the number of up and
down steps are individually preserved. Whether this local
conservation law changes the scaling dimensions on large
scales is the central issue we address here.  From the KPZ
perspective, your initial guess would
probably be ``no", and from the lattice gas perspective ``yes". Our
results presented below confirm the ``no", but
in rather subtle manner, the universality class is ``(KPZ)$^2$"
instead of simple KPZ.

The AHR model has been widely studied recently, with as  focus the
structure of its stationary state \cite{Arndt98, Arndt99, Rajew00, 
Kafri02, Schutz03, Kafri03}.
We are not aware of any previous dynamic scaling analysis.
The stationary state shows strong clustering, as function of
decreasing passing versus free
hopping probability, $r/p$. Stretches of ``empty" road are followed
by high density clusters.
These are mixtures of $S=+1$ and $S=-1$ particles.
We will identify the amount of mixing with the quasi-particles
and the cluster size with the screening length.

The passing of $+$ and $-$ particles resembles collisions.
The ratio $r/p$ controls the duration of the
collision (the softness of the balls).
This passing delay creates queuing and is the origin of the
clustering. The full AHR model includes
a reverse-passing probability  $t$, (($-+)\to(+-)$; particles switching
position in the direction opposite to the electric field).
That enhances the clustering even more. We limit ourselves here to
the $t=0$ version of the model.

The clustering extends over such large length scales, in  specific
ranges of $r/p$ and $t/p$,
that the possibility of a phase transition into a macroscopic
clustered stationary state has been the major issue \cite{Arndt98, Arndt99}.
Macroscopic cluster condensation (infinite sized
clusters) have been shown to be impossible
using the analytic matrix product ansatz \cite{Rajew00} and also 
using an approximate
mapping onto the so-called zero range process \cite{Kafri02}. The 
cluster size remains always
finite, but the maximum value can be far beyond  all computation 
capabilities \cite{Rajew00}.

\begin{figure}[b]
\begin{center}
\includegraphics*[width=0.6\columnwidth, angle=0] {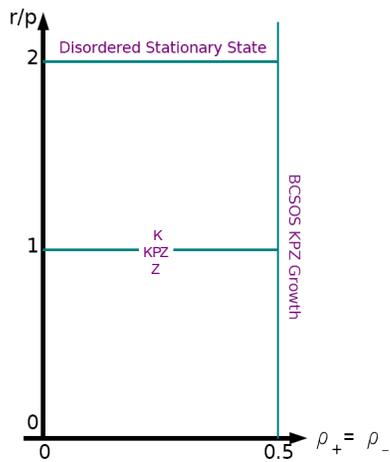}
\end{center}
\caption{\label{phasediagram-fig} Phase diagram for the AHR model
as function of $r/p$ and (conserved) average density $\rho=\rho_+=\rho_-$.
}
\end{figure}

\section{The Phase Diagram}
\label{phasediagram}

Fig.\ref{phasediagram-fig} shows the phase diagram of the
$t=0$ AHR model  as function of $r/p$ and  (conserved) global average
density $\rho=\rho_+=\rho_-$.
It contains three special lines:
$r/p=1$, $r/p=2$, and $\rho=0.5$, respectively.

Along the $\rho=1/2$ line all sites are fully occupied and the
process reduces to the singe-species ASEP.
 From the  perspective of the AHR process as modeling two coupled
conserved degrees of freedom,
momentum and density, the density sector freezes out, leaving only
the Burgers equation.
The  $\rho=1/2$ limit is therefore anomalous,
and this line is not the proper backbone of the phase diagram.
The dynamic scaling exponent is equal to $z=3/2$ along this line, but
that does not need to extend to $\rho<0.5$.

The $r/p=1$ line and the interpretation of the AHR process in terms
of two coupled Burgers equations
form the true backbone of the phase diagram.
At $r=p$, the process reduces to a single-species ASEP in two different ways.
If the $+$ particles choose to be blind to the difference between an
empty site and a $-$ particle,
they see  at $r=p$ no difference between a free hop and a passing
event, and thus experience pure single species
ASEP scaling. The same is true in the projected subspace where $-$
particles are blind to the difference between
empty sites and $+$ particles. These subspaces are not perpendicular
and the process does not factorize
into two  independent ASEP processes. Correlations exist between the
$+$ and $-$ particles, resulting in clustering.
We will study  this numerically in the next section
and find that at large length scales the process
factorizes after all, into  $\mbox{(KPZ)}^2$.

At $r\neq p$ the particles can still pretend to be blind to the other 
species, but
then experience updates where the hopping probability inexplicably 
changes from $p$ to $r$.
These events are random, but not uncorrelated.
For $r<p$ the clustering increases and for $r>p$ decreases.
The line $r=2p$ is special;  there the clustering vanishes 
accidentally altogether.

\section{Dynamic Perfect Screening at $r=p$}
\label{rispscreening}

\begin{figure}
\begin{center}
\includegraphics[width=8cm]{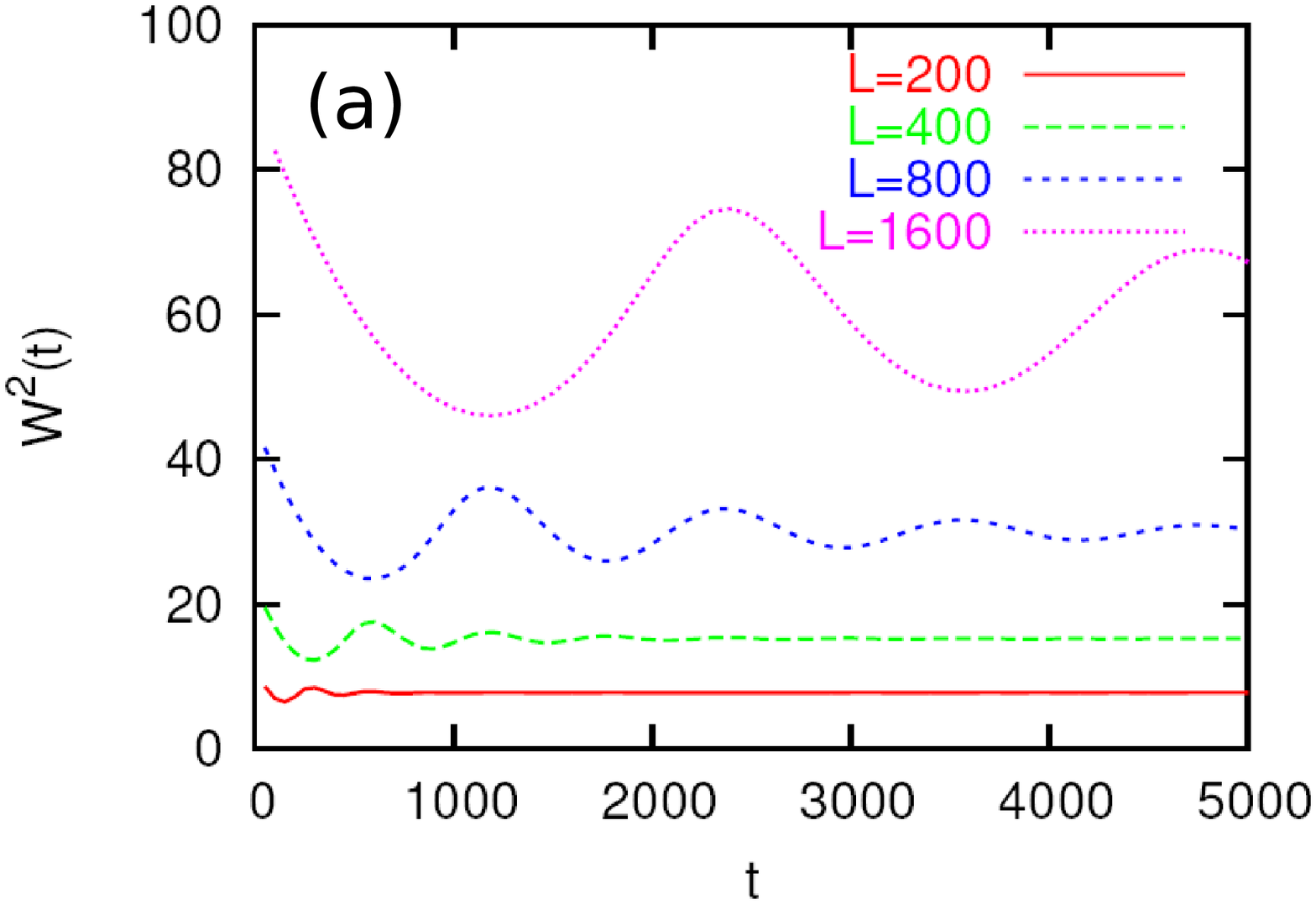}
\includegraphics[width=8cm]{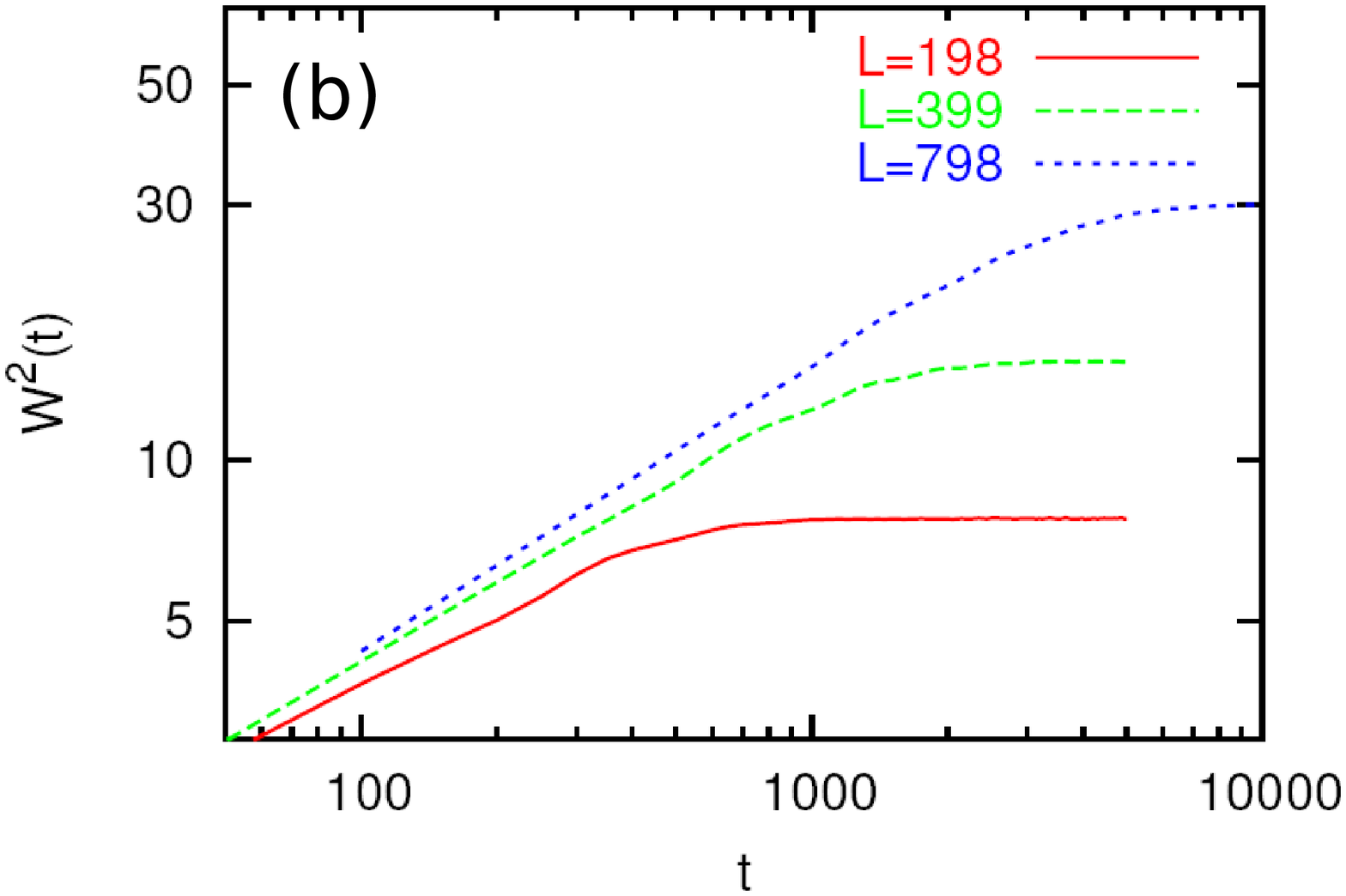}
\caption{\label{width-osc} The evolution of interface widths with
uncorrelated disordered initial states with $r=p=1.0$  and $\rho=0.25$
(a) and flat initial states with $r=p=1.0$ and $\rho=1/3$
(b)  for different system sizes $L$. }
\end{center}
\end{figure}

Our investigation of dynamic scaling in the AHR model started with  an attempt
to measure the dynamic critical exponent $z$ in the conventional manner,
i.e., from the time evolution of the interface width starting from a
flat or a random initial state.
Recall that the AHR model is a RSOS type KPZ growth model with
conserved number of up and down steps.
It turns out that this interface width oscillates in time while
evolving toward the stationary state,
as illustrated  in Fig.\ref{width-osc}.

The flat initial state evolves roughly in accordance with conventional
scaling, i.e., as $W\sim t^{\beta}$, with $\beta =\alpha/z$,
at intermediate times and saturating at $W\sim L^\alpha$ (with
$\alpha$ the stationary state roughness exponent),
but the oscillations on top of this behavior are too strong to
accurately determine $\beta$. These oscillations reflect the additional
conservation law, and are tied to
traveling wave packets propagating in opposite directions and
meeting again after traveling around the lattice ring.

For the resolution of this problem we turn our attention towards
these wave-packets themselves, by monitoring
the manner they spread in time. This is achieved in terms of the
two-point correlators
\begin{eqnarray}
{\cal G}_{+-}(x,t) &=& \langle n_+(0) n_-(x) \rangle - \langle n_+(0) \rangle
\langle n_-(x) \rangle
\end{eqnarray}
and ${\cal G}_{++}$ and ${\cal G}_{--}$, where $n_{\pm}(x)$ is the 
number operator for $\pm$  particles at site $x$
and at time $t$.
The perfect screening phenomenon in the stationary
state emerged while we tested this novel method.
In  this section we first present and discuss perfect
screening and then  present the numerical analysis of the dynamic
exponent, both at $r=p$.

\subsection{Stationary State Correlation Functions}
\begin{figure}[b]
\includegraphics[angle=-90,width=8cm]{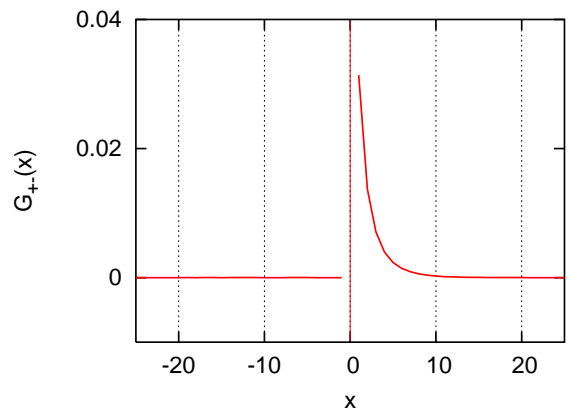}
\caption{\label{p10r10-corxppm-stt} ${\cal G}_{+-}(x)$ in the
stationary state with
$r=p$, $\rho=0.25$, and $L=800$. }
\end{figure}

\begin{figure}[t]
\includegraphics[angle=-90,width=8cm]{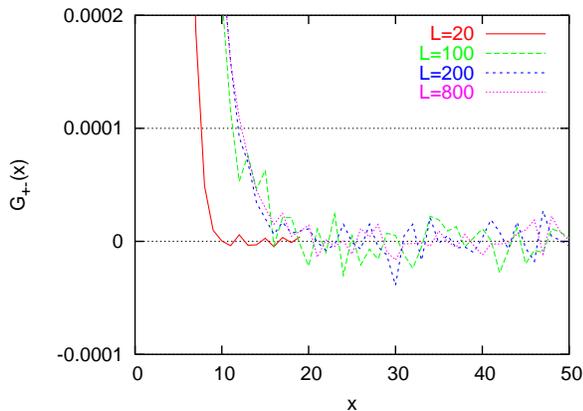}
\caption{\label{corxpm-L} Offsets of ${\cal G}_{+-}(x)$ in stationary
states for
$L=20,100,200,800$ with $\rho=0.25$. }
\end{figure}

In the stationary state, the correlation function
\begin{equation}
{\cal G}_{+-}(x) = \langle n_+(0) n_-(x) \rangle - \rho^2.
\end{equation}
decays exponentially toward zero.
Figs.\ref{p10r10-corxppm-stt}-\ref{corxpm-L} illustrate this, using 
MC simulations
for periodic boundary conditions for small rings, $L\leq 800$.
The correlation function decays exponentially for $x>0$ and is zero for $x<0$.
Correlations are absent at $x<0$,
because after passing, $-$ and  $+$ particles hop away
from each other, and (at $r=p$) do not communicate
with each other anymore.

The correlation length is  rather short in
Fig.\ref{p10r10-corxppm-stt}, $\xi\simeq 5$, but increases with
density
along the $r=p$ line.
The most significant aspect is not the correlation length, but the
absence of any finite size scaling offset
${\cal G}_{+-}(x)\sim B/L$ for $x \gg\xi$ and $x<0$.

The absence of this offset is quite surprising.
It indicates a ``perfect screening" localization type phenomenon in
the fluctuations.
To appreciate this, consider the two-point correlation in a random
disordered state,
like the single species ASEP stationary state.
The ${\cal G}_{++}$ and ${\cal G}_{--}$ correlators in our model have
exactly that form at
$r=p$ because each couples only to one of the two projected single
species ASEP subspaces.
Such correlators are $\delta$-functions (with negative ${\cal G}(0)/L$ offsets)
because periodic boundary conditions
imply rigorous global conservation of the total number of particles,
and impose the condition that
the total area underneath ${\cal G}(x)$ is exactly equal to zero.

Another way of viewing this starts by realizing that
${\cal G}_{+-}(x)/\rho$ can be interpreted as the probability to find
a $-$ particle at distance $x$ from a tagged
$+$ particle at site $x_0$. The tagging removes an amount of
probability $\rho$ from $x_0$ corresponding
to the (untagged) probability of finding a $-$ particle at $x_0$. This
amount is redistributed over the chain.
In general, we would expect that
part of this expelled probability remains localized near
$x_0$, represented in ${\cal G}_{+-}$
by the area underneath the exponential; and that the
remainder is  distributed uniformly over the chain in delocalized
form, represented by a uniform $B/L$ type finite size offset in 
${\cal G}_{+-}$.
For uncorrelated $\delta$-function  type correlations all of it is delocalized,
such that  $B=\rho^2$.
Our numerical simulations, see Fig.\ref{corxpm-L}, put
a bound on the delocalized  amplitude; e.g.,
$|B|/L< 8 \times 10^{-6}$ at $L=800$  for $\rho=0.25$.
The delocalized fraction is zero within the MC noise.

So surprisingly, in our process all the excluded probability is localized,
such that
\begin{equation}
{\cal G}_{+-}(0) = - \sum_{y=1}^{y=a} {\cal G}_{+-}(y)
\end{equation}
for all $\xi\ll a \ll L$.
A person riding on top of a specific $+$ particle and wearing glasses
that filter out the $-$ particles,
observes a perfect single species ASEP in terms of the $+$ particles.
Without glasses she notices however an excess of $-$ particles in
front of her. This cloud  of size $\xi$
has on average an excess mass equal to $\rho$.

\subsection{Factorization from Perfect Screening}

The above  perfect screening implies that the AHR process at $r=p$ behaves at
coarse grained scales as two decoupled single species Burgers equations.
This factorization is easily recognized in the interface growth representation.
Recall that the $+$ particles represent up-steps and the $-$
particles down steps, and that the number of both  are
conserved. Perfect screening means factorization into two decoupled
KPZ interface growth
processes at length scales $x \gg \xi$ (one where down-steps are
being ignored and the other where the up-steps are ignored).

The interface width $W(x)$ of the full model over a section of the
interface of length $x$  can be expressed
in terms of the two-point correlators as
\begin{eqnarray}
W(x)^2 &\equiv& \big\langle (h(x)-h(0))^2 \big\rangle \nonumber\\
      &=& \Big\langle \Big(\sum_{y=0}^x (-n_+(y)+n_-(y))\Big)^2 
\Big\rangle \nonumber\\
      &=& \Big\langle \Big(\sum_{y=0}^x (n_+-\rho_+) - \sum_{z=0}^x
(n_--\rho_-)\Big)^2 \Big\rangle \nonumber\\
      &=& \sum_{y,z=0}^x \Big[{\cal G}_{++}(y-z)+{\cal G}_{--}(y-z) \nonumber\\
	   &&~~~~- {\cal G}_{+-}(y-z) - {\cal G}_{-+}(y-z)\Big].
\end{eqnarray}
${\cal G}_{++}$ and ${\cal G}_{--}$ are $\delta$-functions at $r=p$
and their finite size offsets are absent in the thermodynamic limit
$L\to
\infty$,
\begin{eqnarray}
W(x)^2     &=& x (\rho_+-\rho_+^2)+x (\rho_--\rho_-^2) + 2x (\rho^2 -
A_{+-}).\nonumber
\end{eqnarray}
Moreover, at length scales much larger than the screening length, $x \gg \xi$,
the cross-correlator area $A_{+-} \equiv \sum_{y=1}^x {\cal
G}_{+-}(y)$ reduces to $A_{+-}=\rho^2$ by perfect screening, such that
the ${\cal G}_{+-}$ contributions vanish completely,
\begin{eqnarray}
W(x)^2
      &=& x(\rho_+-\rho_+^2) + x(\rho_-- \rho_-^2) \nonumber\\
      &=& W(x;+)^2 + W(x;-)^2.
\end{eqnarray}
The  square of the full interface width is thus equal to the sum of
the squared interface widths in the two projected subspaces
at $x \gg \xi$.
The two coupled Burgers equations behave independently at length 
scales $x \gg \xi$.

\subsection{Dynamic Exponent from ${\cal G}_{+-}(x,t)$}

\begin{figure}
\includegraphics[angle=-90,width=8cm]{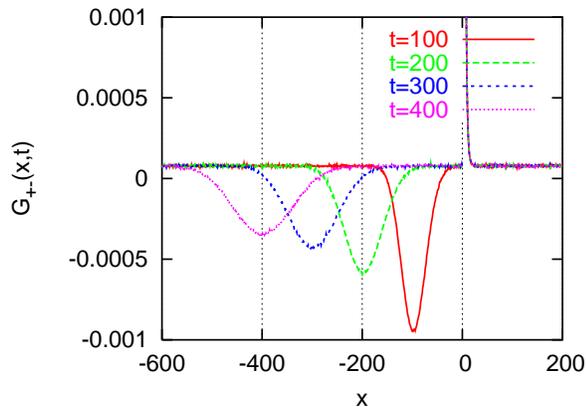}
\caption{\label{p10r10-corxpm} ${\cal G}_{+-}(x,t)$ at series of time
$t=$100, 200, 300, 400 with $r=p=1.0$, $\rho=0.25$, and $L=800$. The group
velocity is equal to  $v_g = 2 p(1-2 \rho) =1$.}
\end{figure}

Fig.\ref{p10r10-corxpm} shows the time evolution of the ${\cal G}_{+-}$
correlation function
starting from an initial uncorrelated disordered state (a 
$\delta$-function with a
finite size off-set).
The build-up of the cluster of $-$ particles in front of the tagged 
$+$ particle
requires only a short time span $\tau_0$.
The build-up of this surplus is mirrored by the build-up of a
depletion layer behind the $+$ particle (particle numbers  are 
locally conserved).
After the screening cloud at $x>0$ is fully established, $t>\tau_0$,
the depletion packet detaches from $x=0$ and travels to the left.
This traveling wave packet belongs to one of the two projected single species
ASEP subspaces and
therefore should spread in time with KPZ dynamic exponent $z=3/2$ as
$w \sim t^{1/z}$.
The Gaussian form
\[
\delta {\cal G}_{+-}(x,t)  \sim t^{-1/z}\exp\Big[
-\frac{(x-v_gt)^2}{Dt^{2/z}}\Big],
\]
fits the wave packet very well \cite{Leven} except at times close to
$t=\tau_0=\xi/v_g$ where it is slightly skewed.
The packet's group velocity, $v_g$, follows the expected value $v_g =
2p (1-2\rho)$,
i.e., twice the group velocity of fluctuations in the $-$ or $+$
sector single species ASEP. ($2v_g$  is the relative
velocity of fluctuations in the $+$ and $-$ sectors
respectively, propagating in opposite directions.)
The traveling depletion wave packet moves around the ring while
broadening. It collides after one period
with the screening cloud. They split-off again.
This keeps repeating itself, until the broadening has
spread all over the ring and
cancels out against the global finite scaling offset of the initial state.

\begin{figure}[th]
\begin{center}
\mbox{\includegraphics[angle=-90,width=8cm]{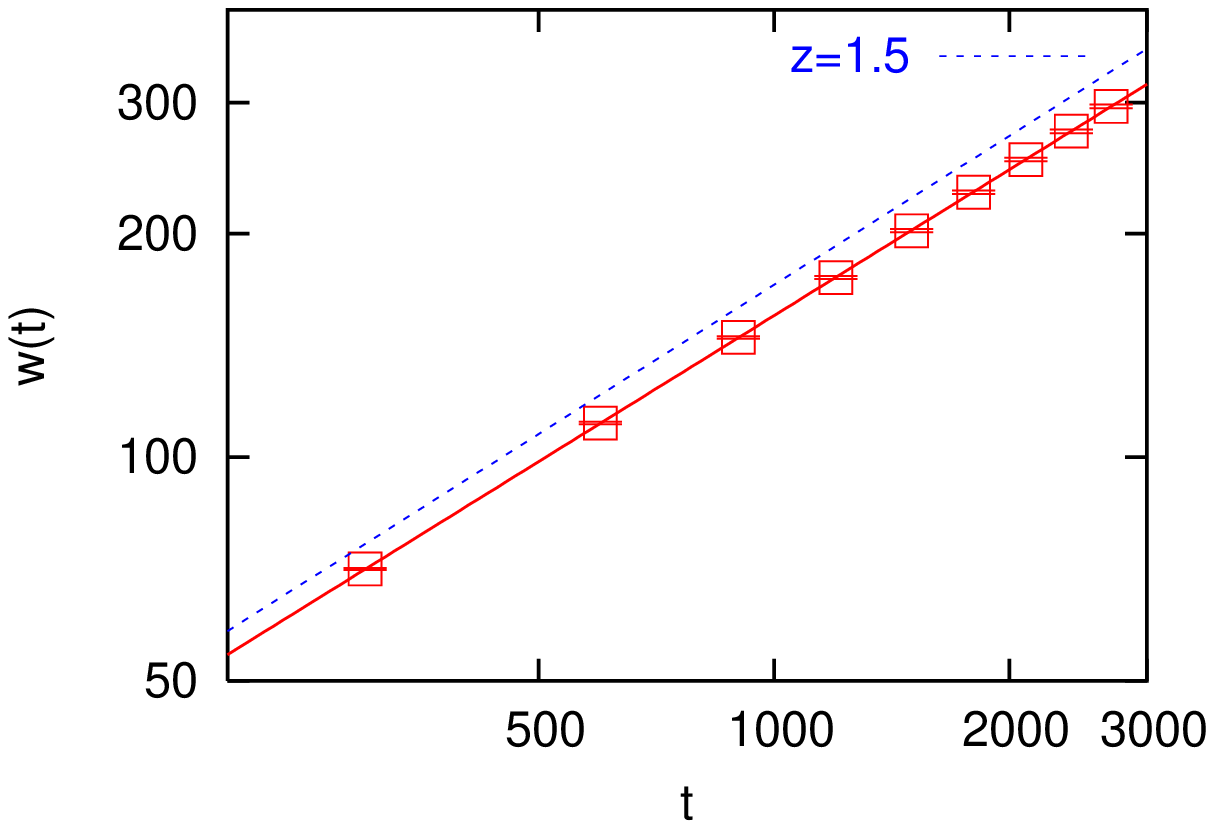}}
\mbox{\hspace{-0.5cm}\includegraphics[angle=-90,width=8.7cm]{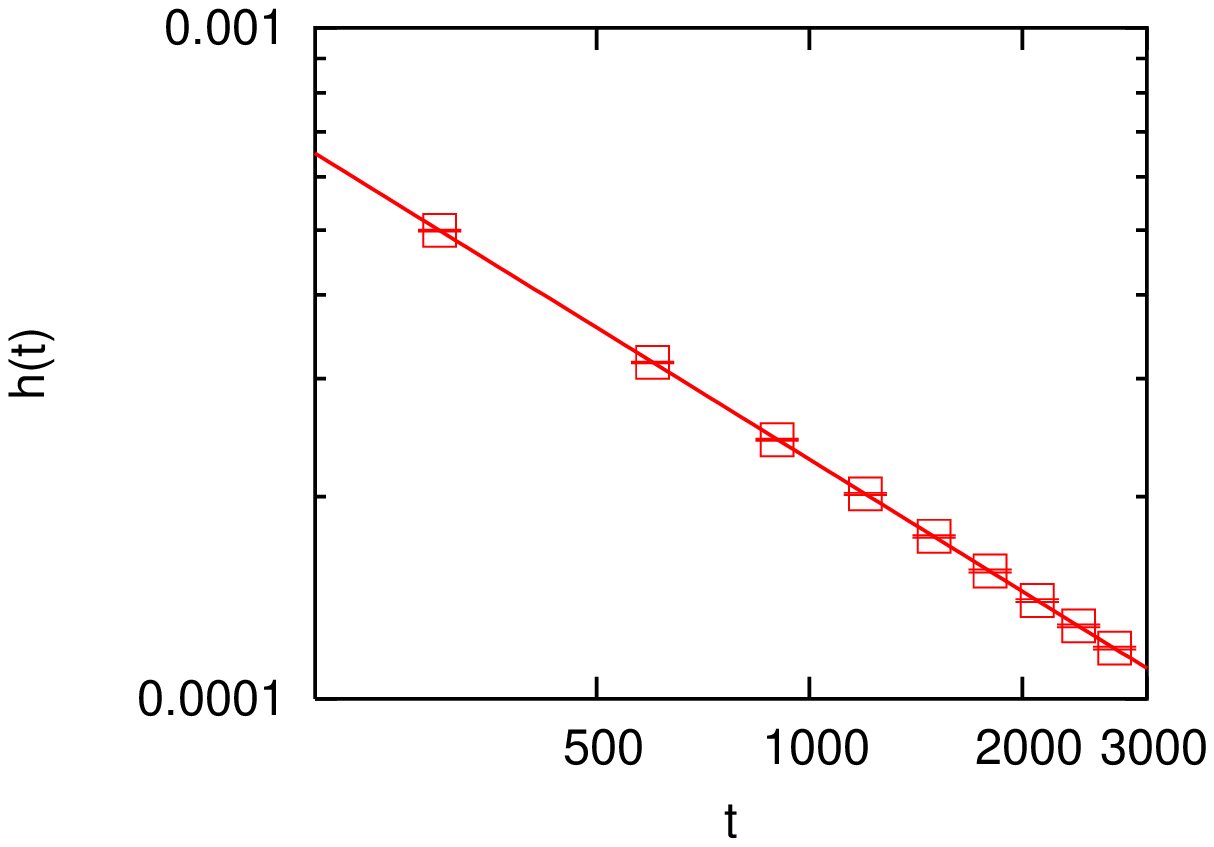}}
\caption{\label{p10r10-fit} Widths and heights of the
wave-packets in ${\cal G}_{+-}(x,t)$ at series of time, $t$, for $p=r=1.0$ and
$L=3200$. A line corresponding to $z=1.5$ is drawn above the data
points in the upper figure.}
\end{center}
\end{figure}

\begin{figure}[th]
\begin{center}
\includegraphics[angle=-90,width=8cm]{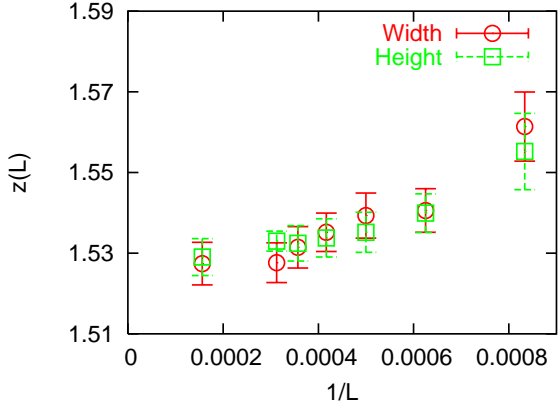}
\caption{\label{p10r10-z} The estimates of the dynamic
exponent, $z$, for finite system sizes $L=$1200, 1600, 2000, 2400,
2800, 3200, 6400.  Analyses of finite size correction to the scaling
shows the estimate of dynamic exponent is equal to $1.53[2]$.}
\end{center}
\end{figure}

Fig.\ref{p10r10-fit}  shows the time evolution of the width of the wave packet
$w$ and its  height $h$. They obey  power laws: $w \sim t^{1/z}$ and
$h \sim t^{-1/z}$.
 From these, we obtain estimates for the dynamic exponent $z$, and
Fig.\ref{p10r10-z} shows the finite size scaling behavior of these estimates.
They converge to  $z=1.53[2]$, consistent with the expected KPZ value
$z=3/2$. This confirms that this novel way for determining $z$ works well.

\section{Dynamic Screening at $r \neq p$}
\label{rnotpscreening}

The correlation functions ${\cal G}_{+-}$ and ${\cal G}_{++}$ take
more intricate shapes
away from the $r=p$ line.  Remarkably, as we
will discuss next, this variety of shapes convert back into the
simple shapes of $r=p$
using a quasi-particle representation. We discovered this
numerically, as presented in this section,
and then proved it analytically, as presented in the next section.
The properties at the $r=p$ line, perfect screening between particles
of opposite charge and uncorrelated
disordered stationary state statistics in the two projected
subspaces, extend thus to
all $r/p$ in terms of quasi-particles, and the final conclusion from
this is that
the process factorizes into (KPZ)$^2$ everywhere for all $r/p$.

\subsection{Stationary State Correlation Functions}

\begin{figure}
\mbox{
\includegraphics[width=4.3cm]{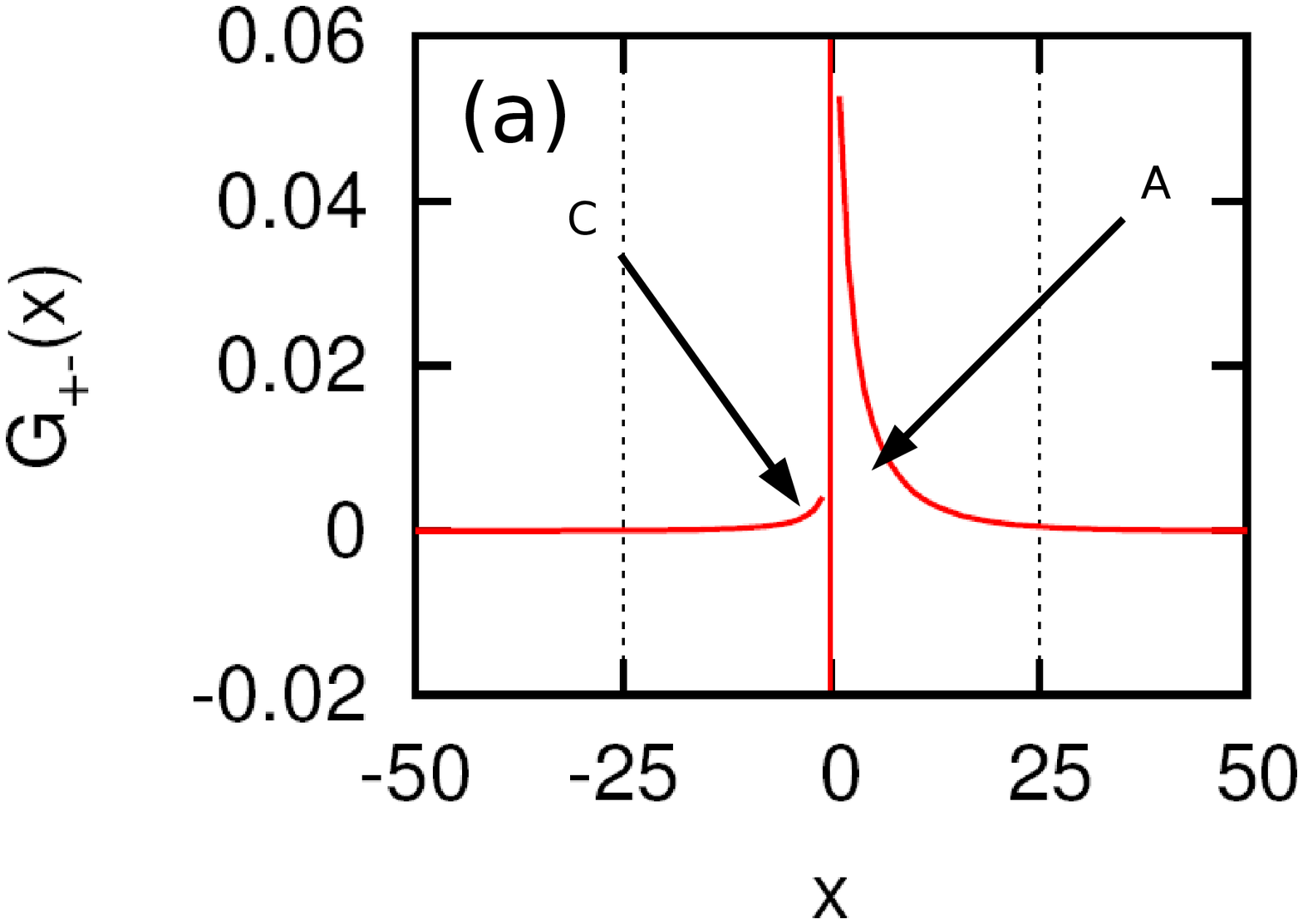}
\includegraphics[width=4.3cm]{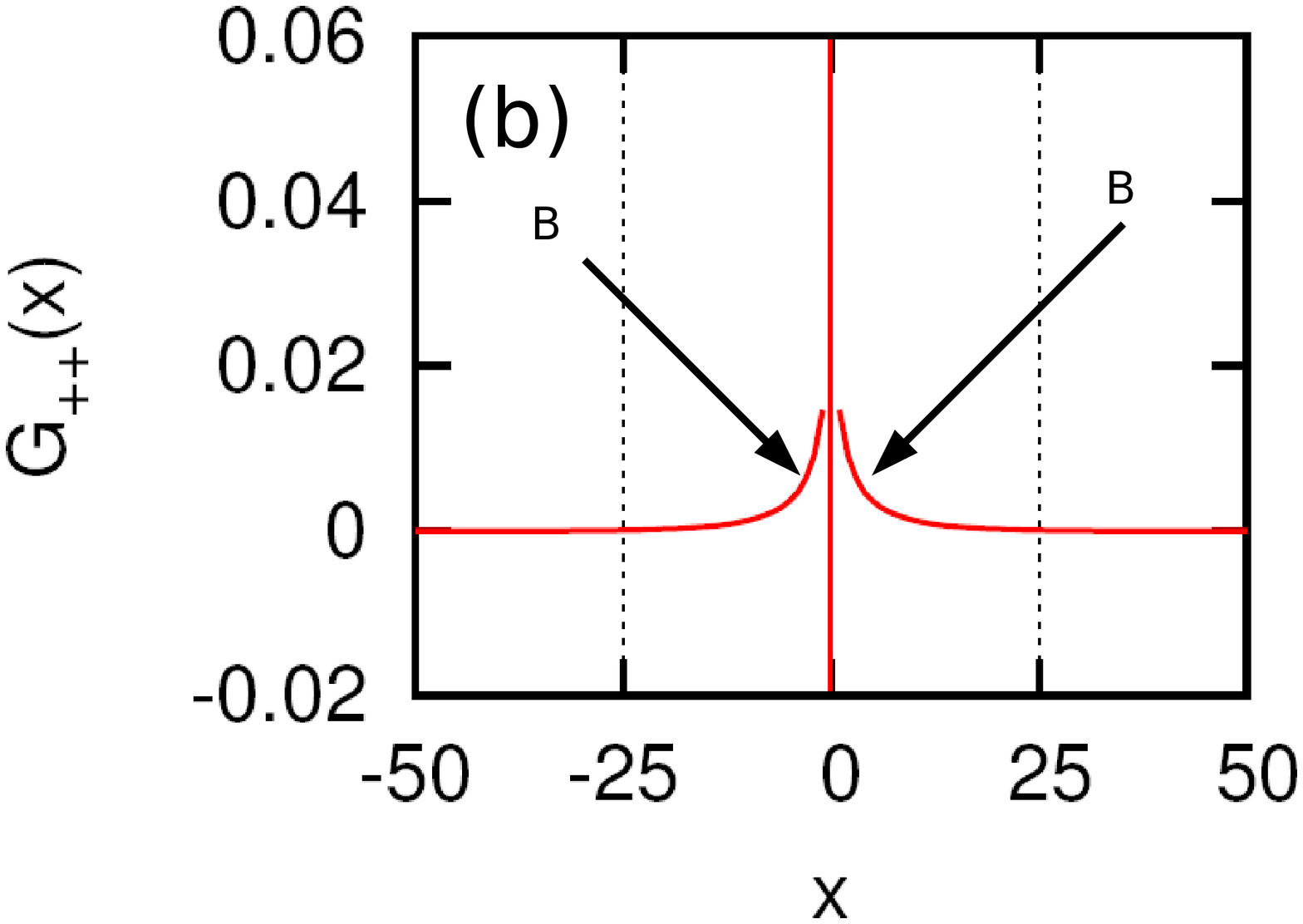}
}
\mbox{
\includegraphics[width=4.3cm]{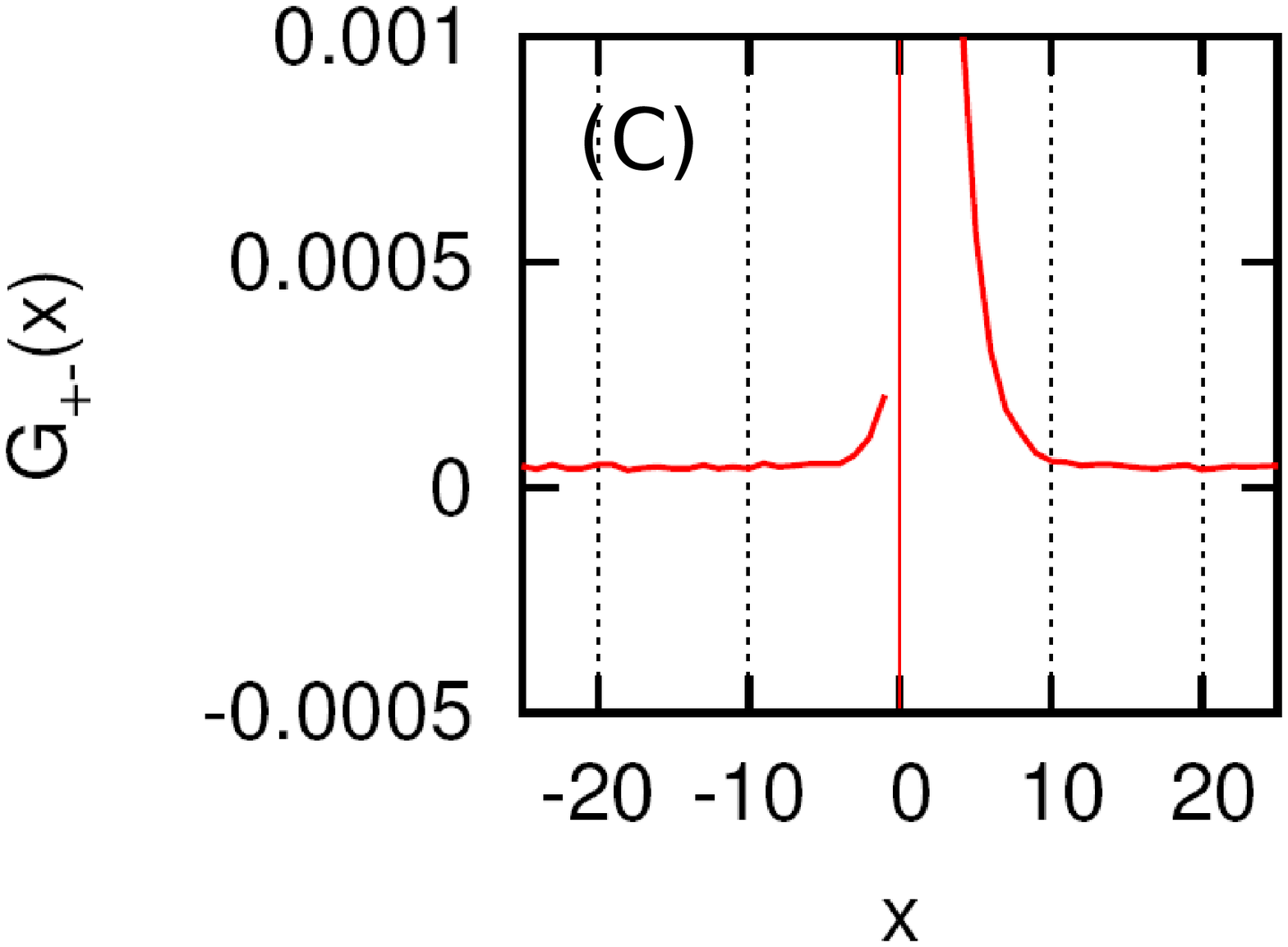}
\includegraphics[width=4.3cm]{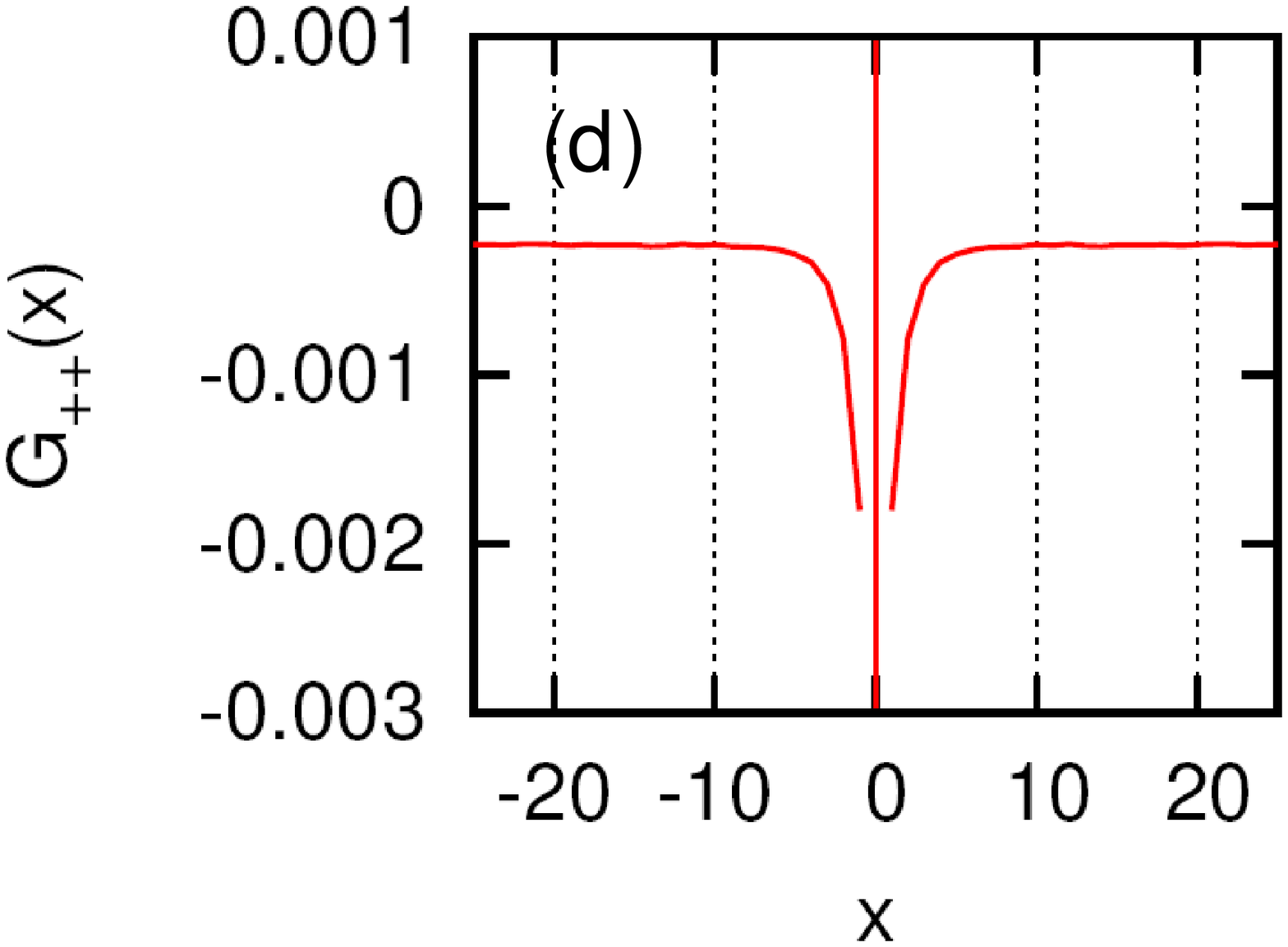}
}
\mbox{
\includegraphics[width=4.3cm]{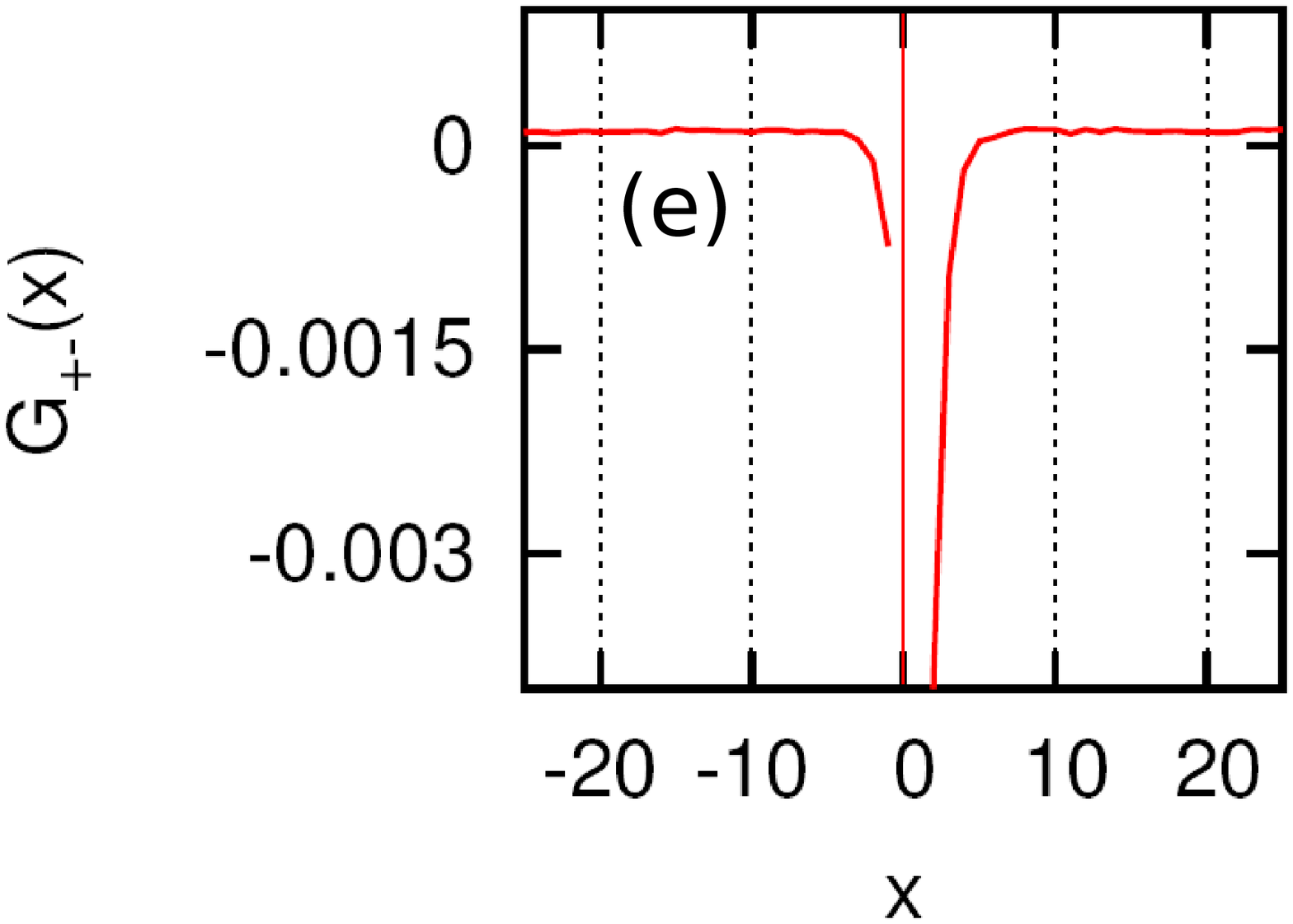}
\includegraphics[width=4.3cm]{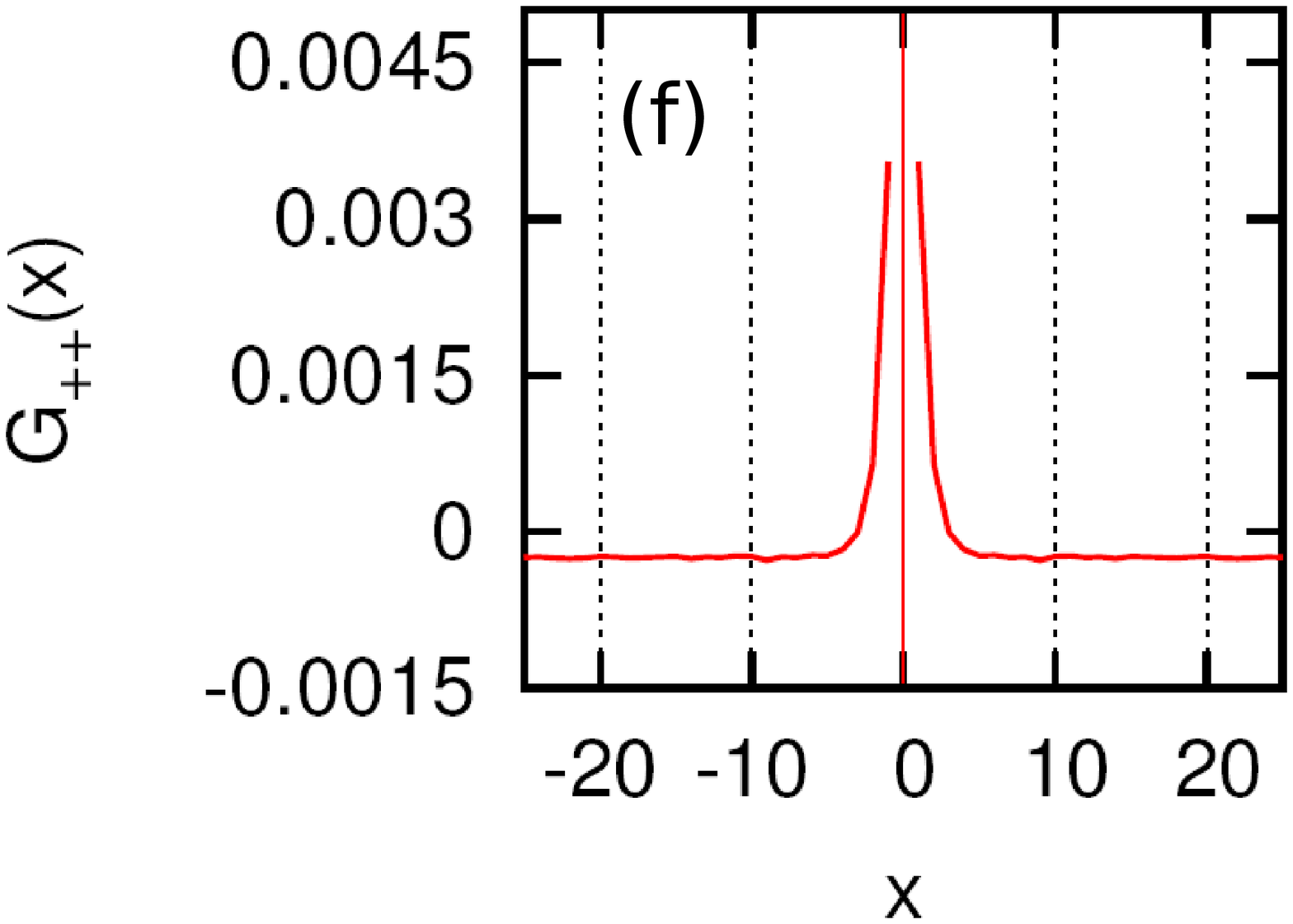}
}
\caption{\label{r-neq-p-cor} Stationary correlation fictions,
${\cal G}_{+-}(x)$ (left column) and ${\cal G}_{++}(x)$ (right column), for
$p=1.0$, $r=0.5$ and $\rho=0.25$ for $L=3200$ ((a) and (b)),
$p=0.7$, $r=1.0$ and $\rho=0.25$ for $L=800$ ((c) and (d)), and
$p=0.3$, $r=0.9$ and $\rho=0.25$ for $L=800$  ((e) and (f)).}
\end{figure}

Fig.\ref{r-neq-p-cor} shows the ${\cal G}_{+-}$ and ${\cal G}_{++}$
correlators for various values of $r/p$.
Compared to the $r=p$ shapes, ${\cal G}_{+-}$ develops correlations
at $x<0$, and ${\cal G}_{++}={\cal G}_{--}$ changes from a $\delta$-function
into a symmetric correlated shape.
This can be  explained qualitatively as follows.
At $r\neq p$, the $+$ and $-$ particles can not choose to be blind
with respect to each other anymore.
Additional correlations build-up compared to the $r=p$
baseline behavior:

At  $r<p$ the passing versus hopping rate is reduced. The screening
cloud at $x>0$ in ${\cal G}_{+-}$ therefore grows (the clustering is stronger).
This enhanced ${\cal G}_{+-}$ screening cloud at $x>0$,
results in short range correlations between alike particles as well;
${\cal G}_{++}(x)={\cal G}_{--}(x)$ develops positive tails.
This is a second order effect. Those $++$ particle correlations in turn induce
positive correlations in ${\cal G}_{+-}(x)$ for $x<0$.
This is a third order effect, and thus an order of magnitude further down.

At $r> p$ the passing rate is enhanced with respect to the $r=p$
baseline behavior.
The $x>0$ screening cloud in ${\cal G}_{+-}(x)$ is thus smaller than at $r=p$.
The correlations in ${\cal G}_{++}={\cal G}_{--}$ are indeed negative,
and represent  a reduced probability to find alike particles near each other.
This reduced probability makes it less likely
to find $+$ particles behind the tagged $+$ particle, at $x<0$.
If those $+$ particles  had been there, they would carry smaller 
screening clouds in front of them.
Their absence therefore creates still positive correlations  between $-$
particles at $x<0$ and the tagged $+$ one.

At $r=2p$ the stationary state is fully disordered \cite{Arndt98},
the clustering vanishes and  all correlation functions reduce there
to $\delta$-functions.
At $r>2p$ the correlation tails re-emerge, but with opposite signs.

\subsection{Dynamic Exponents from ${\cal G}_{+-} (x,t)$ and ${\cal
G}_{++} (x,t)$}\label{sec-r-neq-p-z}
\begin{figure}
\begin{center}
\includegraphics[width=8cm]{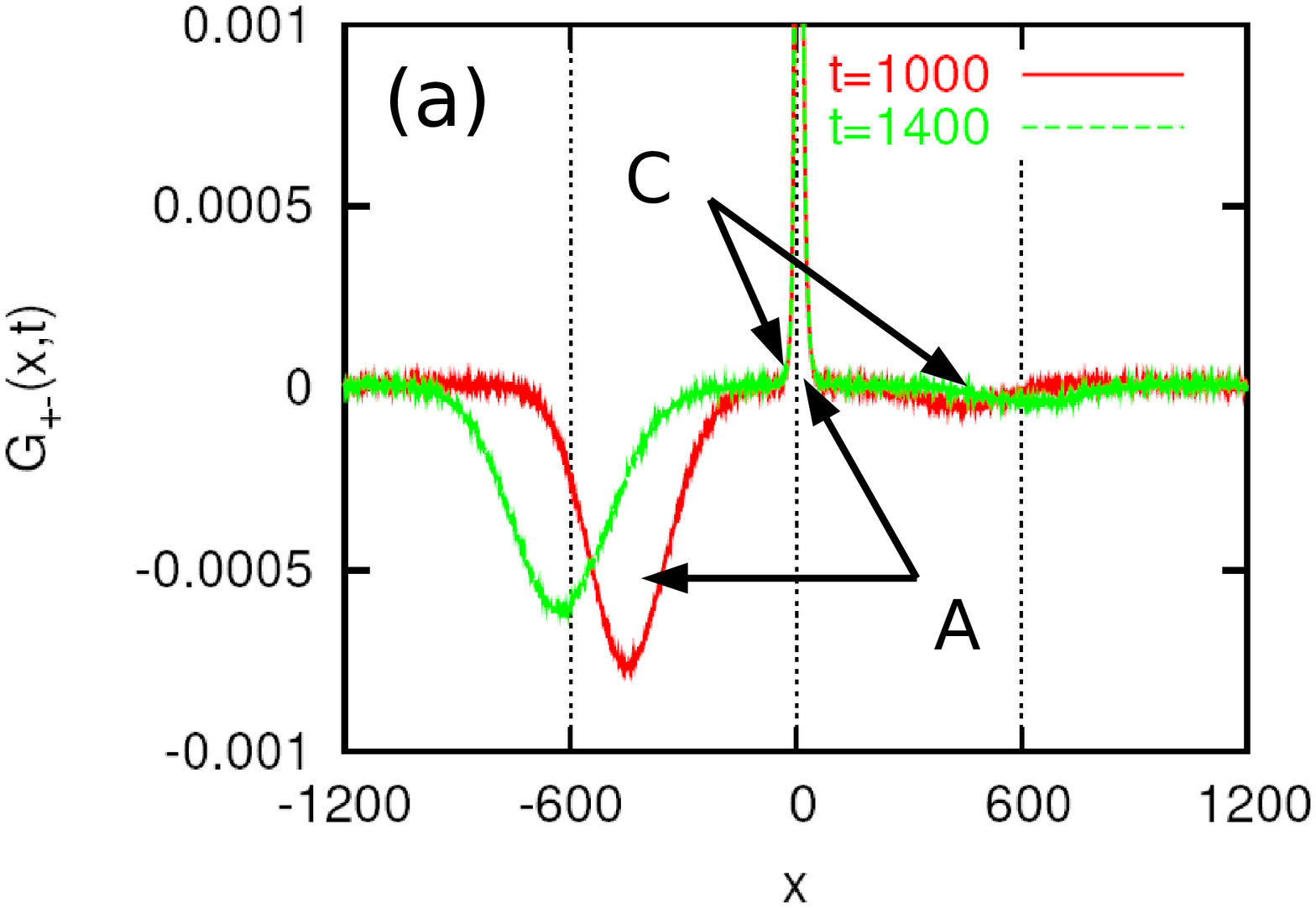}
\includegraphics[width=8cm]{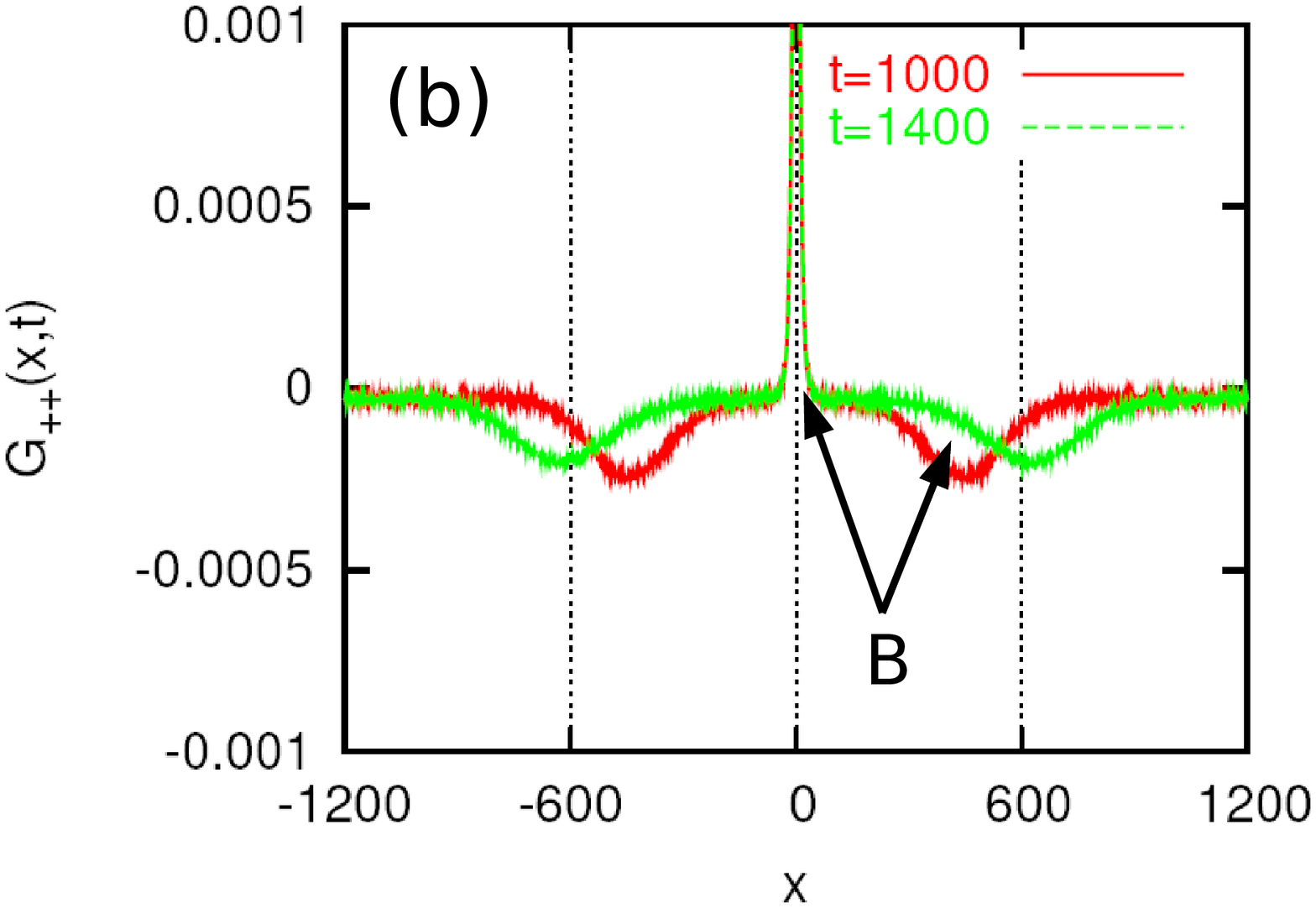}
\includegraphics[width=8cm]{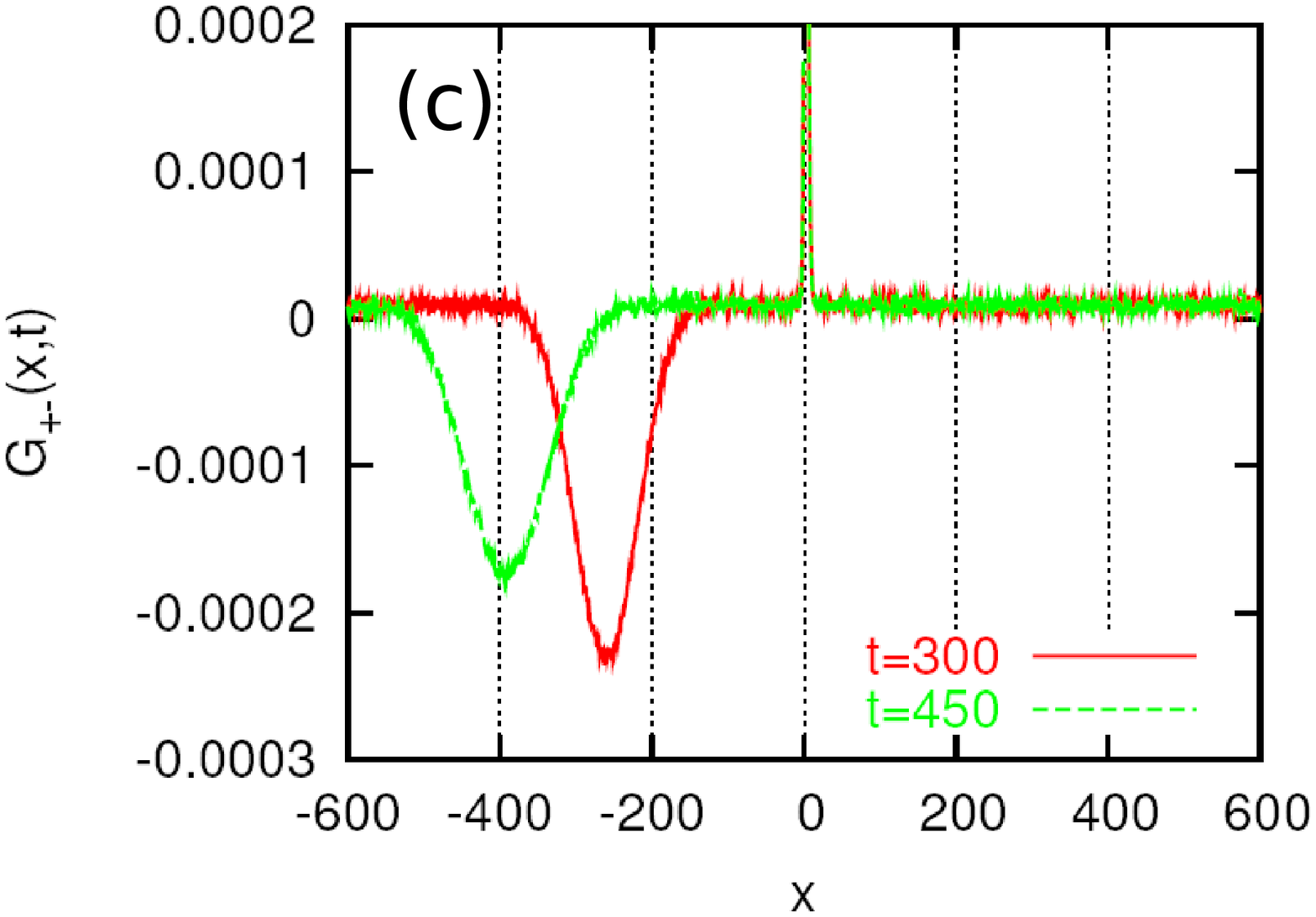}
\includegraphics[width=8cm]{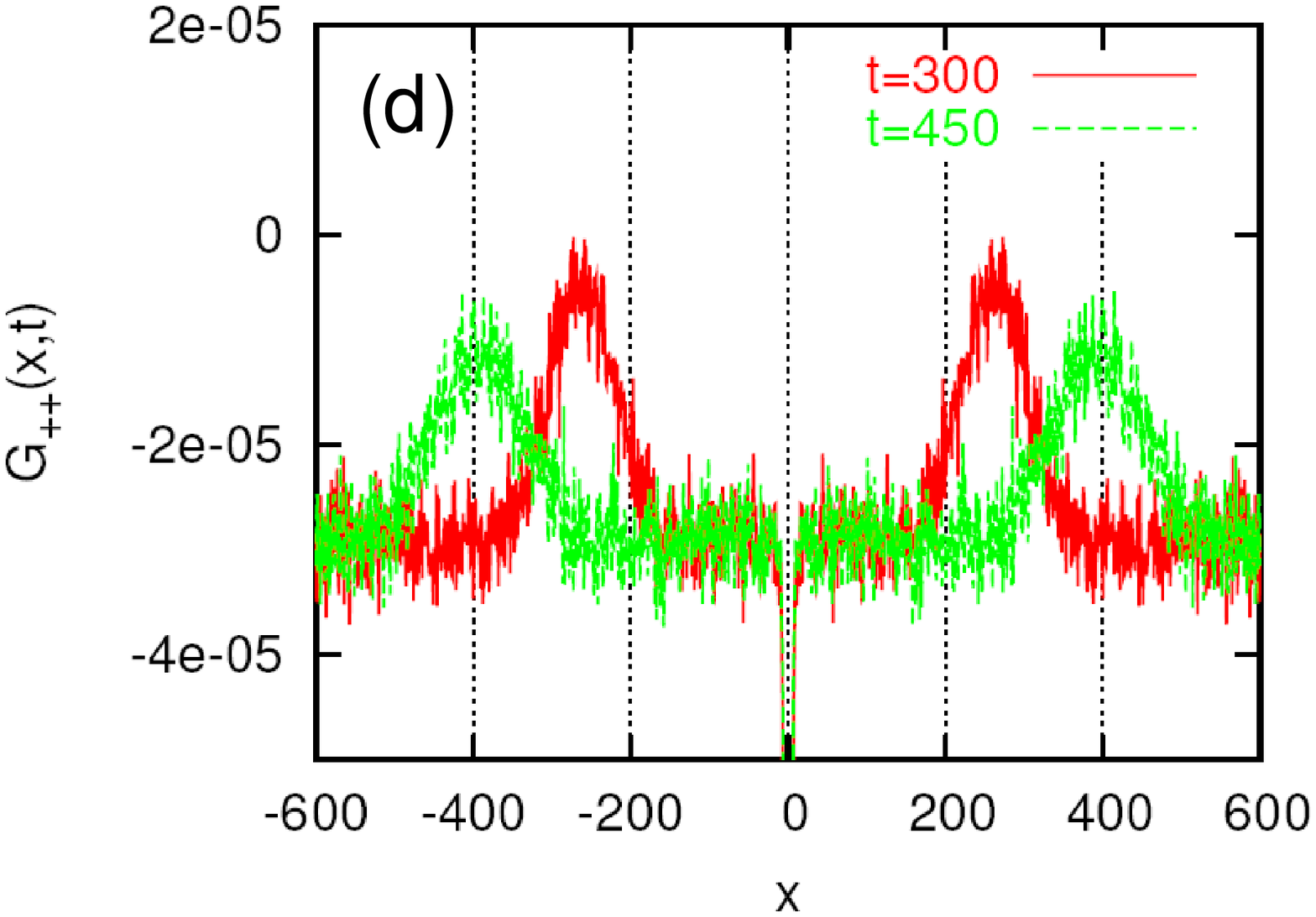}
\caption{\label{r-neq-p-corxppm} (a) Correlation function
between $+$ and $-$ at $t=1000$ and $1400$ with $p=1.0$, $r=0.5$,
and $L=6400$.  The initial state is random disordered. (b) The
corresponding correlation function between $+$ and $+$. (c)
${\cal G}_{+-}(x,t)$ for $t=300$, $450$ with $p=0.7$, $r=1.0$, and
$L=6400$.  (d) The corresponding ${\cal G}_{++}(x,t)$.}
\end{center}
\end{figure}

We examine the temporal evolution of ${\cal G}_{+-}(x,t)$ and ${\cal
G}_{++}(x,t)$
using MC simulations, just as we did in the $r=p$ case. The initial
states are prepared to be uncorrelated
and disordered. As shown in Fig.\ref{r-neq-p-corxppm}, two
wave-packets appear, with different amplitudes, but moving in
opposite directions with the same speed.
The wave-packets in ${\cal G}_{+-}$ are strongly coupled to those in
${\cal G}_{++}$.
These traveling clouds are generated by the same type of mechanism as
the one at $r=p$,
i.e., the result of the rather fast build-up of the screening clouds near
the tagged particle,
reflected by the short distance correlations in  the stationary state.
Both traveling  clouds are  mixtures of $+$ and $-$ particles, with
non-zero projections in both
${\cal G}_{++}$ and ${\cal G}_{--}$.

\begin{figure}
\includegraphics[angle=-90,width=8cm]{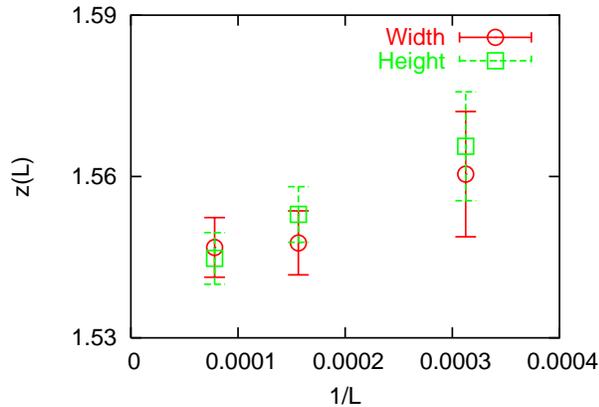}
\caption{\label{p10r5-z}  The estimates of dynamic
exponent, $z$, for different system sizes at $p=1.0$, $r=0.5$,
and  $\rho=0.25$.}
\end{figure}

\begin{figure}
\includegraphics[angle=-90,width=8cm]{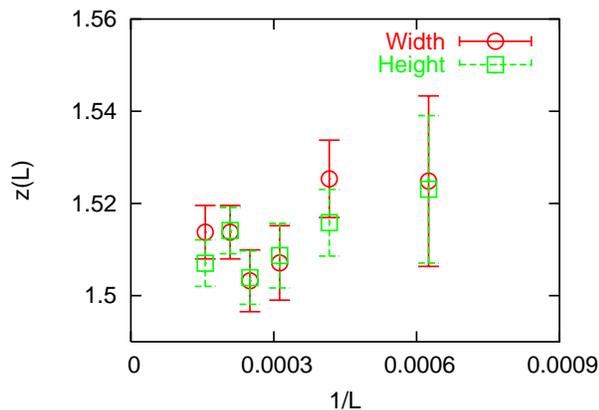}
\caption{\label{p7r10-z}  The estimates of dynamic
exponent, $z$, for different system sizes at $p=0.7$, $r=1.0$, and
$\rho=0.25$.}
\end{figure}

Once the clouds are detached from $x=0$, they move independently of
each other in opposite directions; just as at $r=p$.
The process factorizes again.
But there is no a priori reason why these mixed traveling clouds at $r\neq 
p$ should spread as in pure KPZ.
However, they do.
In our  MC simulations they  spread, e.g., at  $r=0.5$, $p=1.0$,
and $\rho=0.25$, with  $z=1.54[2]$
and at $r=0.7$, $p=1.0$, and $\rho=0.25$ with  $z=1.51[2]$.
Fig.\ref{p10r5-z} and \ref{p7r10-z} shows strong finite size corrections to
the scaling in the dynamic exponents, but the limiting behavior is clear.

\begin{figure}
\begin{center}
\includegraphics[width=7.5cm]{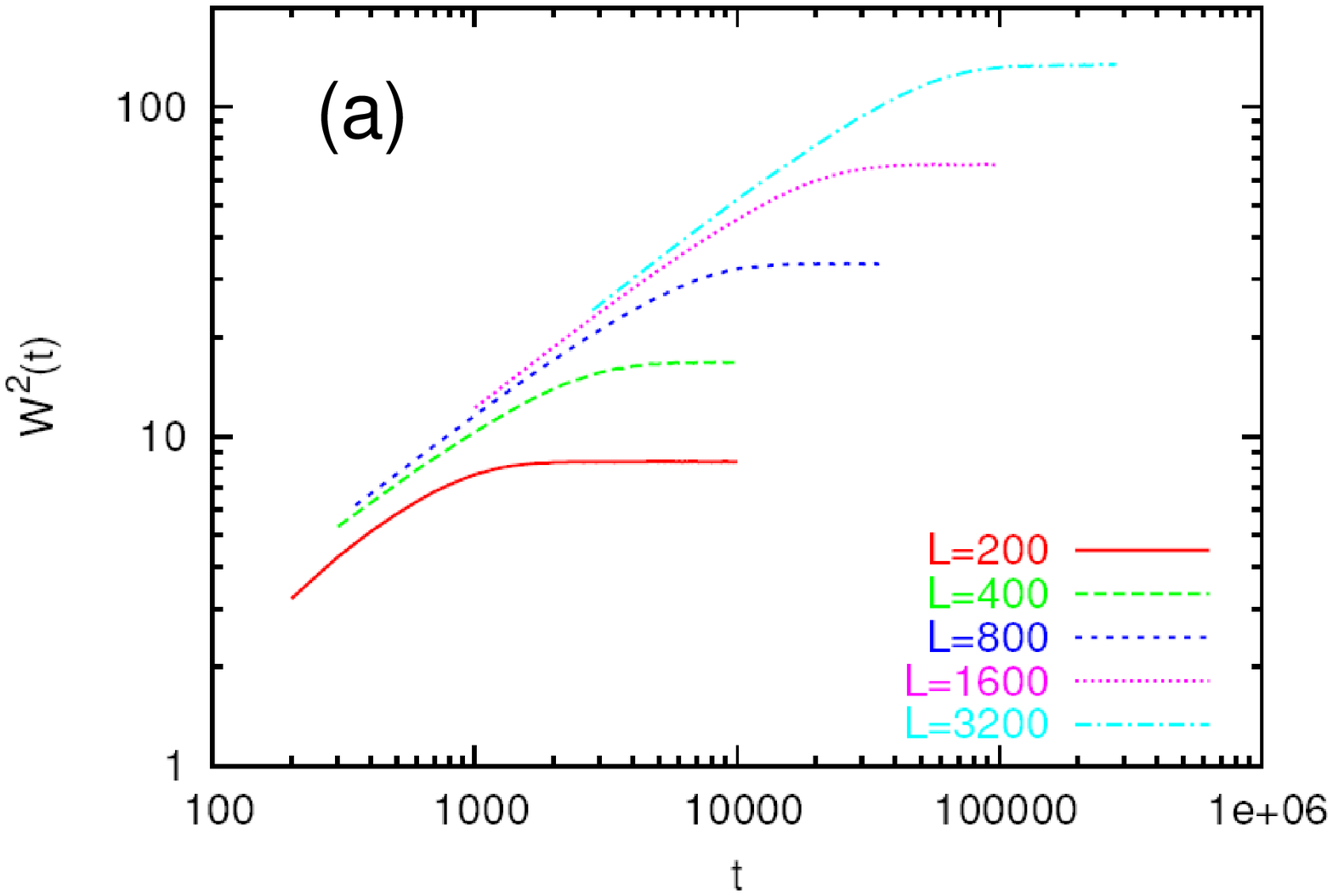}
\includegraphics[width=7.5cm]{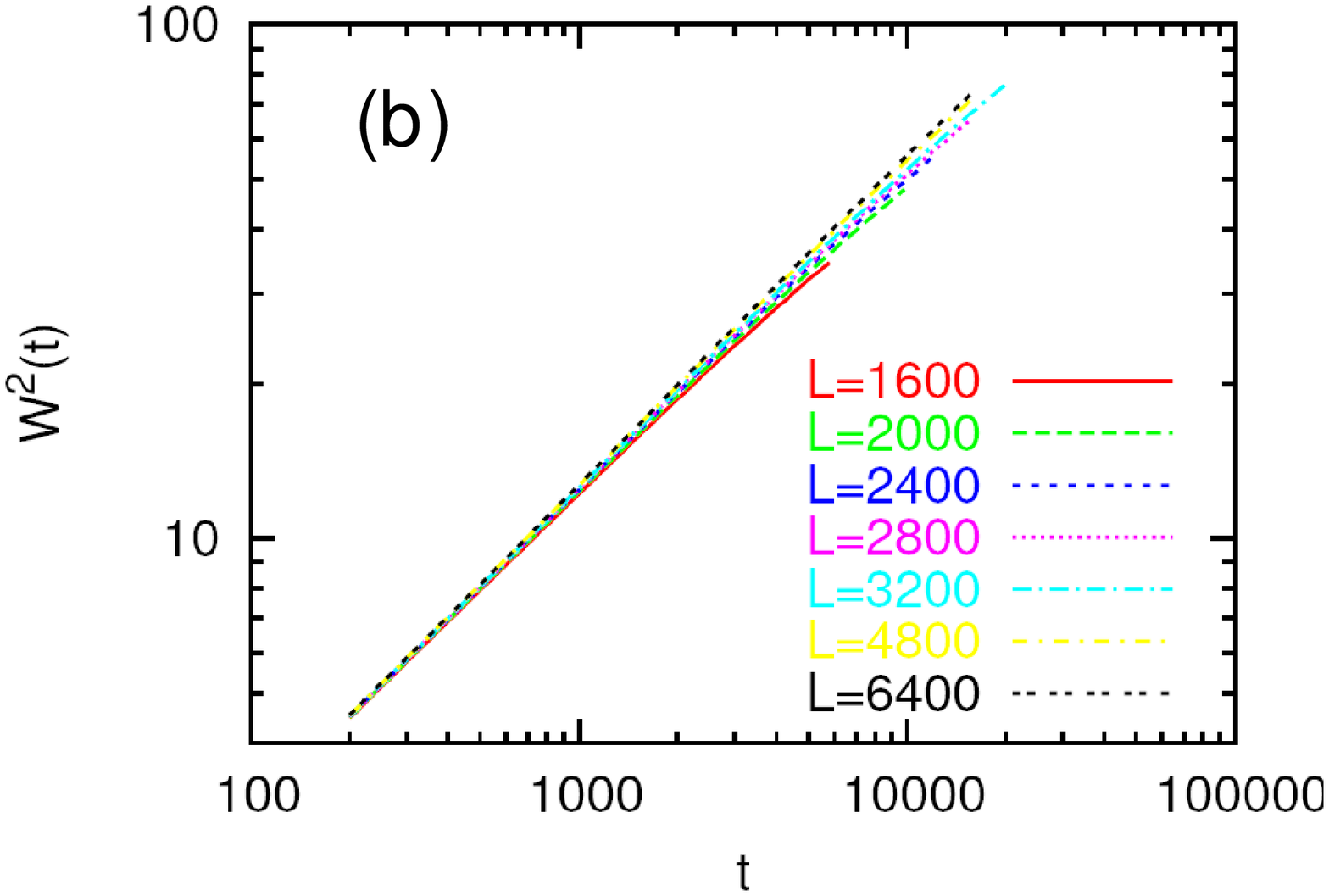}
\caption{\label{r-eq-2p-width}  The temporal evolution of the
interface widths starting
from initial flat interfaces for $p=0.5$, $r=1.0$, and $\rho=0.25$.
(a) The evolution of the
widths for different system sizes only shows oscillations for $t<200$.  (b)
The dynamic exponent is
estimated by measuring the slopes of log-log plot of the interface
width v.s. time.}
\end{center}
\end{figure}

\begin{figure}
\includegraphics[angle=-90,width=8cm]{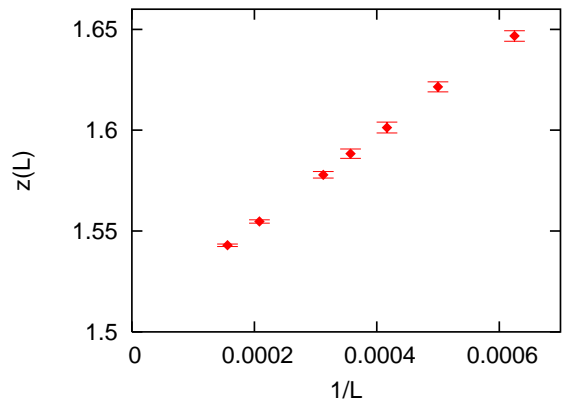}
\caption{\label{r-eq-2p-z} The estimate of dynamic
exponent, $z$, for different system sizes at $p=.5$, $r=1.0$ and $\rho=0.25$.}
\end{figure}

Moreover, at $r=2p$,  the stationary state is totally uncorrelated
and disordered (and the temporal evolutions of the correlation
functions therefore do not involve traveling wave packets).
We can apply  the conventional method to estimate the dynamic
exponent.  The temporal
evolution of the interface widths (see Fig.(\ref{r-eq-2p-width}) and
(\ref{r-eq-2p-z})), yields $z=1.51[1]$.

\subsection{Perfect Screening in the Quasi-Particle Representation}

If indeed the dynamic exponent retains (KPZ)$^2$ type value at $r\neq p$ as
suggested by the above numerical results, then
there might be a quasi-particle description in which the process
factorizes at large length scales and
in which the fluctuations are perfectly screened, just as at $r=p$.
We found such a description, first numerically as described here, and
then rigorously
analytically in the following section.
This implies the process obeys  $(\mbox{KPZ})^2$ scaling everywhere.
In terms of quasi-particles the dynamic process
fully factorizes into  two KPZ processes at large coarse-grained scales.

Consider the stationary state correlation functions in
Fig.\ref{r-neq-p-cor} and \ref{r-neq-p-corxppm}:
  the correlation functions decay to $1/L$
type finite size scaling off-sets.
The area $A$ underneath  ${\cal G}_{+-}$ for $x>0$, the area $B$
underneath ${\cal G}_{++}$ at $x>0$ (equal to the same for $x<0$)
and the area $C$ underneath ${\cal G}_{+-}$ for $x<0$ obey
empirically the relation
$B/A=C/B$, for all $r\neq p$, typically with a numerical accuracy,
$1-B^2/AC = 0.01 \%$. (The areas are measured with respect to the offsets.)

This special balance in the areas relates to
a specific amount of mixing between  $+$ and $-$ particles in the
clouds, and suggests (a much stronger property) the existence of
a quasi-particle representation,
\begin{equation}
n_p = \alpha n_+ + \beta n_- \quad n_m=\beta n_++ \alpha n_-,
\label{quasip}
\end{equation}
with  $n_{\pm}$ the  number operator for $+$ and $-$ particles and
\begin{equation}
\label{mixing}
\frac{\beta}{\alpha}= \frac{B}{A}=\frac{C}{B}
\end{equation}
in which the correlation functions ${\cal G}_{pp}(x)={\cal
G}_{mm}(x)$  and ${\cal G}_{pm}(x)$,
defined as
\begin{equation}
{\cal G}_{\nu\mu}(x) \equiv \langle n_\nu(0) n_\mu(x) \rangle - \langle
n_\nu(0) \rangle \langle n_\mu(x) \rangle,
\end{equation}
with $\nu,\mu=p,m$ reduce to the same shapes as the particle
correlators  at $r/p=1$
(where ${\cal G}_{pp}(x)$ is a $\delta$-function and ${\cal G}_{pm}$ has
only one tail and shows perfect screening
between quasi-particles of opposite charges).

\begin{figure}[!ht]
\mbox{\hspace{-1.5cm}\includegraphics[angle=0,width=10.5cm]{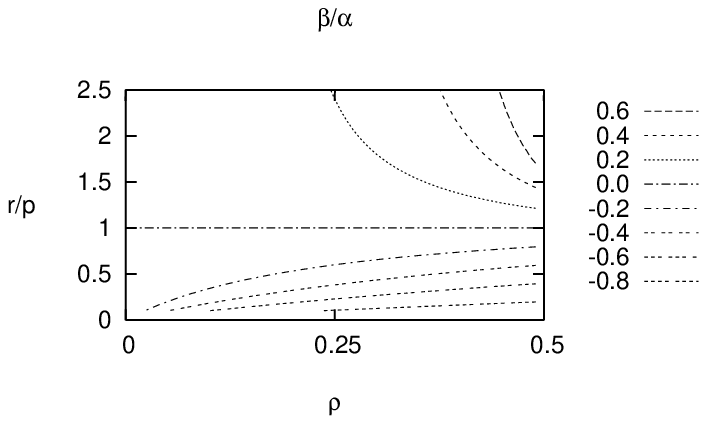}}
\caption{\label{mixing-strength} Contour plot of $\beta/\alpha$ in a
parameter space of $r/p$ and $\rho$ (See Eq.(\ref{b-ov-a})).}
\end{figure}

The mixing ratio $R=\beta/\alpha$ varies from $R=\beta/\alpha=0$ at $r=p$ (with
$n_p=n_+$ and $n_m=n_-$);
to $R=\beta/\alpha$=1 when $n_p=n_m$, and to $R=\beta/\alpha=-1$ when
$n_p=-n_m$.
Fig.\ref{mixing-strength} shows lines of $R$ from our
analytic expression  in section \ref{quasi-r-neq-p}.
Our numerical results are completely consistent with this.
The mixing strength increases with density  $\rho$, and becomes
indeterminate at the line $\rho=0.5$, where all sites are fully occupied.
At $r/p=2$ the stationary state is totally disordered,
but  $R$ does not vanish since $+$ and $-$
remain  strongly correlated dynamically \cite{Schutz04}.
Both $\alpha$ and $\beta$ go to zero and change sign across the $r/p=2$ line.

\section{Perfect Screening and the Matrix Product Stationary State Structure}
\label{MPA}

In this section, we prove analytically the perfect screening of the
(quasi-particles) pair correlators, using the matrix product ansatz (MPA)
structure of the stationary state.
The proof applies to all $r/p$, but for clarity we split-up the discussion.
First we review briefly the general properties of MPA stationary states.
Next, we present the proof at $r=p$, and finally generalize it to all
$r/p$ in terms of quasi-particles.

\subsection{MPA Type Stationary States}

Stationary states of stochastic dynamic processes are typically very complex
with intricate long-range effective interactions between the  degrees
of freedom
(when writing the stationary state  in terms of effective
Gibbs-Boltzmann factors).
The long-range aspect is important; 1D driven stochastic processes can undergo
non-trivial phase transitions, while 1D equilibrium degrees of
freedom with short-range interactions can not.
MPA states are linked to equilibrium distributions and therefore lack
long-range correlations.

MPA stationary states are of the form
\cite{Arndt98, Arndt99, Rajew00}:
\begin{equation}
P_s(\{\tau_i\}) ={1\over Z} Tr[ G_{\tau_1} G_{\tau_2} \cdots ]
\label{MPA-eq}
\end{equation}
with in our case $\tau_i=+,0,-$.
This structure resembles closely the transfer matrix formulation of
partition functions in
one dimensional (1D) equilibrium statistical mechanics.
Consider for example a one dimensional Ising model, with
spin $S=\pm 1$ degrees of freedom  at sites $i+\frac{1}{2}$, that interact as
\begin{equation}
E=\sum_i K(\tau_i) S_{i+\frac{1}{2}}S_{i-\frac{1}{2}}
\end{equation}
and with  a degree of freedom $\tau_i=0,\pm 1$ on every bond $i$
such that bond energy $K(\tau_{i})$  can have three  distinct values.
In that case, the $G_{\tau_i}$ are $2\times 2$ transfer matrices and
the stationary state probability
for the yet unrelated  stochastic dynamic process is the Ising equilibrium
partition function for a given  $\{\tau_i\}$ configuration,
\begin{equation}
P_s(\{\tau_i\})= \sum_{\{S\}} e^{-E(\{\tau,S\})}.
\label{partition}
\end{equation}
The normalization factor
\begin{eqnarray}
{\cal Z}&=&\sum_{\{\tau_i\},*}  P_s(\{\tau_i\})
\label{partition1}
\end{eqnarray}
is the canonical partition function of the  annealed random bond 1D
Ising model.
The stochastic driven non-equilibrium dynamics typically imposes
constraints on the
$\tau_i$ degrees of freedom. In
our dynamic process the number of each species of particle,
$\tau=\pm$, is conserved independently.
This is denoted by $*$ in eq.(\ref{partition1}).

The $\tau$ variables do not couple to each other directly in
eq.(\ref{partition});
all correlations between $\tau$ degrees of freedom are mediated by
the Ising field $S_x$.
The search for a possible MPA structure of the stationary state is
therefore  the search for the existence of
a representation in which all correlations between the original
$\tau$ degrees of freedom are
carried by a new auxiliary field and expressed as short-range
interactions between those new degrees of freedom.
Those auxiliary degrees of freedom can take any form, not just Ising
spins, because the rank of
the $G$ matrices and their symmetries
can be arbitrary. For example in our case, the rank will be infinite,
and the auxiliary field can be interpreted as
(integer valued) interface type degrees of freedom, denoted as
$n_x=0,\pm1,\pm2, \cdots$.

The transfer matrix product structure, eq.(\ref{MPA-eq}),
implies that those auxiliary degrees of freedom  interact by nearest
neighbor interactions only.
This is actually unfortunate, because in short-ranged 1D equilibrium
systems, like eq.(\ref{partition}),
spontaneous broken symmetries and phase transitions are impossible.
Therefore, master equations with  MPA stationary states have at best
dynamic phase transitions with trivial scaling properties (associated
with an abrupt change in the $G$-representation).
For example, MPA  representations of directed percolation or directed
Ising type processes can not exist,
because both are believed to have
transitions with complex scaling dimensions.
Still, the MPA method has been proved to be a powerful tool,
its algebraic structure is very elegant, and a surprisingly large
class of 1D stochastic dynamic processes
have a MPA type stationary state.

Boundary conditions play an important role.
Eq.(\ref{partition1}) is a canonical partition function, where the
number  of $+$ and $-$ particles are each conserved.
Consider instead the generating function
\begin{eqnarray}
{\cal Z}
&=&\sum_{\{\tau_i\}} z_-^{N_-} z_+^{N_+} P_s(\{\tau_i\}) \nonumber\\
&=& Tr[(z_+ G_+ +z_- G_- +G_0)^L] =Tr[M^L]
\label{grandpartition}
\end{eqnarray}
with
\begin{eqnarray}
M = z_+ G_+ + z_- G_- +G_0.
\label{transfer}
\end{eqnarray}
This would be the grand canonical partition function of, e.g., the
above annealed random bond 1D Ising model
in case of periodic boundary conditions.
$z_\pm$ are the fugacities of the $\tau_\pm$ particles.
The equivalence  between the ensembles in the thermodynamic limit is
ensured in the equilibrium interpretation,
where the details of how the particle reservoirs couple to the system
does not have to be addressed.
This is different in the interpretation of the MPA as the stationary
state of a driven stochastic process.
Dynamic processes are very sensitive to boundary conditions.
For example, a process with open boundary conditions and reservoirs
at the edges conserves the number of particles everywhere
inside the bulk, and  behaves very different from
the one where the reservoirs couple directly to every site.
Not surprisingly therefore, the MPA method only applies to the
stationary state;
the introduction of the auxiliary field does not address the
stochastic dynamics, nor the  temporal fluctuations in the stationary state.
For periodic boundary conditions, as in our case,
the grand canonical partition function eq.(\ref{grandpartition})
represents an ensemble of dynamic systems, each with periodic boundary
condition systems,  and fixed values of $N_-$ and $N_+$, weighted
with respect to each other by the fugacity probability factor.
In this sense the ensembles are equivalent in the thermodynamic limit.
In our discussion below we use the grand canonical ensemble.
The correlation functions for $x>0$ are evaluated then as
\begin{eqnarray}
{\cal G}_{_{_{+-}}}\!(x)=\frac{1}{\lambda_B^2} \Big[\langle B|G_+
\left(\frac{M}{\lambda_B}\right)^{x-1} \! G_-|B\rangle
\nonumber\\  -\langle B|G_+|B\rangle\langle B|G_-|B\rangle\Big]
\label{correl}
\end{eqnarray}
in the thermodynamic limit, with
$|B\rangle$ and $\langle B|$ the right and left eigenvectors of the
largest eigenvalue $\lambda_B$
of the operator $M$ defined in eq.(\ref{transfer}). The correlator at $x=0$,
$\langle n_+n_-\rangle$, poses somewhat of a problem. It can not be
expressed as  simple as this  due to the intrinsic
off-diagonal character of the above $G$ operators.
At $r=p$ this is not an issue, because $\langle n_+n_-\rangle=0$.
However, that will not be true anymore for
the quasi-particles at $r\neq p$.

\subsection{Quadratic Algebra}

The first step in identifying whether the stationary state of a
stochastic process might have a MPA structure,
is to insert Eq.(\ref{MPA-eq}) into the master equation.
If lucky, the condition of stationarity can be expressed as simple
algebraic conditions on the $G_\tau$ transfer matrices.
The MPA structure of our process has been studied extensively
recently \cite{Arndt98, Arndt99, Rajew00, Kafri02, Schutz03, Kafri03}.
 From those studies we know that the three $G_\tau$ must obey the quadratic
algebra:
\begin{eqnarray}
r G_+ G_-  &=& -x_- G_+ + x_+ G_-,\nonumber\\
p G_+ G_0  &=& -x_0 G_+ + x_+ G_0,\nonumber\\
p G_0 G_-  &=& -x_- G_0 + x_0 G_-,
\label{quadalgebra}
\end{eqnarray}
with $x_0$ and $x_\pm$ arbitrary yet unspecified  parameters.
These conditions apply to the entire phase diagram, for all $r/p$.
The next step is to find explicit representations of the $G_\tau$
that satisfy eq.(\ref{quadalgebra}),
using the freedom in choice of the parameters $x_i$.
In general the rank of the $G_\tau$ does not close, but remains infinite.
The rank is finite only along special lines in the phase diagram.
Fortunately, for our purposes we do not need closure;
the quadratic algebra structure itself is sufficient to prove perfect
screening.

Our process is invariant under simultaneous inversion in space  $x\to -x$
and of charge  $+\leftrightarrow -$ in the case that the
numbers of $+$ and $-$ particles are balanced.
This suggests we look for a realization of the algebra with operators
satisfying
$G_+ = G_-^T$ and  $G_0 = G_0^T$. This invariance  is valid in the
subspace $x_+=-x_-=r$ and $x_0=0$ \cite{Arndt98},
where the quadratic algebra reduces to
\begin{eqnarray}
G_+ G_-  &=& G_+ + G_-\nonumber\\
G_+ G_0 &=& G_0 G_- = \frac{r}{p} G_0.
\label{rqalgebra}
\end{eqnarray}

\subsection{The $r=p$ Quadratic Algebra}

At $r=p$, the quadratic algebra of Eq.(\ref{rqalgebra}) is easily checked to be
satisfied by the operators \cite{Arndt99}
\begin{eqnarray}
G_+ = I+L_-,~~
G_- = I+L_+,~~
G_0 =  |0\rangle \langle 0|.
\end{eqnarray}
The rank of these matrices is infinite. The auxiliary degrees of freedom are
(positive only) integer valued
``height variables" $n=0, 1,2,\cdots$.
$G_0$ is the projection operator onto the  $n=0$ state, and $L_{\pm}$ are
the raising (lowering) operators, $L_\pm |n\rangle = |n \pm 1\rangle$.

We need to determine the eigenvalues of
the grand canonical transfer matrix,  eq.(\ref{transfer}),
\begin{equation}
M =  zG_+ +zG_- +G_0 = z(2I +L_+ +L_-) +G_0.
\label{M}
\end{equation}
This  matrix has several interpretations. It is the transfer matrix
for the equilibrium
partition function of a one dimensional interface in the presence of
a substrate (all $n<0$ are excluded)
with a short range attractive  potential at $n=0$; like a substrate.
Such an interface layer is thin and non-rough.
It is also the time evolution of a 1D random walker (with $x$ playing
the role of time and $n$ that of space)
in half space, $n\geq 0$ and an on-site attractive interaction at
site $n=0$. Such a random walker is localized.
The latter can be presented also as localization of a single quantum
mechanical particle
hopping on a semi-infinite chain with a $\delta$-function attractive
potential at the first site,
\begin{equation}
H \equiv 2I -L_+ -L_- - \frac{1}{z}G_0.
\label{hamil}
\end{equation}
with  $M=4z(1-\frac{1}{4}H)$.

This simple Hamiltonian has one single bound state and a
continuum spectrum of extended states.
The calculation of the eigen-spectrum is elementary and straight forward.
The eigenstates, $|\lambda \rangle = (\phi_0, \phi_1, \cdots)$,
satisfy the equations:
\begin{eqnarray}
(2-\frac{1}{z}) \phi_0 -  \phi_1 &=& E \phi_0 \nonumber\\
-\phi_{n-1} + 2 \phi_n - \phi_{n+1} &=& E \phi_n, \quad
\mbox{for $n\geq 1$.}
\label{r=p-2}
\end{eqnarray}
Bound states have the generic form
\begin{equation}
\phi_n = \frac{1}{\sqrt{Z_B}} w_b^{n} \quad \mbox{for} \quad n \geq
0. \label{bound-state}
\end{equation}
Substitution in Eq.(\ref{r=p-2}) yields only one bound state,
with energy  $E_B=2-1/w_b-w_b$, such that
\begin{equation}
\lambda_B = \frac{z}{w_b} (1+w_b)^2,
\label{bound-energy}
\end{equation}
and normalization
\begin{equation}
Z_B = \frac{1}{1-w_b^2}.
\end{equation}
$w_b$ is equal to $w_b=z$.

The extended eigenstates are scattered waves, with
\begin{equation}
\phi_0=\frac{A_0(k)}{\sqrt{Z(k)}}
~~;~~
\phi_n = \frac{1}{\sqrt{Z(k)}} \cos\big(kn+\theta_k\big).
\label{scatter}
\end{equation}
The eigenvalue equations at
$n>1$ yield the energy spectrum,
$E(k) =2(1-\cos k)$, with $0 < k < \pi$, such that
\begin{equation}
\lambda(k) = 2z(1+\cos k),
\label{spectrum}
\end{equation}
and those at $n=0,1$ yield the phase shift $\theta_k$,
\begin{equation}
A_0(k)=\cos(\theta_k)=z\cos(\theta_k-k)=\frac{z \cos(\theta_k+k)}{2z-zE_k-1}.
\end{equation}
The normalization factor
\begin{equation}
|\phi_0 |^2 + \sum_{n=1}^D | \phi_n |^2=1
~\to~
Z(k) = |A_0|^2+ \frac{1}{2}D
\label{scatter-norm}
\end{equation}
is proportional to the rank of the matrices $D$,
and thus strictly speaking infinite; $D$ will drop out in our
calculations below.
The $n=0$ component is easily evaluated:
\begin{equation}
\phi_0^2= |\langle 0| k \rangle|^2= \frac{2}{D}\frac{ z^2\sin^2
k}{z^2-2z \cos k+1}.
\label{h-zero}
\end{equation}

\subsection{Perfect Screening at $r=p$} \label{perfect-screening-r-eq-p}

Perfect screening implies that
\begin{equation}
\sum_{x=1}^\infty {\cal G}_{+-}(x) = -{\cal G}_{+-}(0),
\label{perfect-screening-r=p}.
\end{equation}
i.e.,  that the sum over $x>0$ of the correlator Eq.(\ref{correl}),
\begin{equation}
S=\sum_{x=1}^\infty {\cal G}_{+-}(x)
\label{sumG}
\end{equation}
is equal to the right hand side of Eq.(\ref{perfect-screening-r=p}),
\begin{equation}
S=\rho^2 = z^2 \left(\frac{\lambda_p}{\lambda_B}\right)^2,
\label{sumrule-r=p}
\end{equation}
with $\rho=z\lambda_p/\lambda_B$, using that the bound state is also
an eigenstate of $G_+$, $G_+|B
\rangle=\lambda_p|B\rangle=(1+w_b)|B\rangle$.  In our specific case the density
is simply equal to $\rho=z/(1+z)$, but we like to
keep the derivation as generic as possible.

We need to demonstrate that this sum rule is valid in the
thermodynamic limit, and track carefully any terms that scale
as system size $L$. For example, as discussed already in detail
above, the sum rule is trivially true for periodic
boundary conditions, but then does not imply perfect screening,
because any unscreened surplus can be spread out over
the entire lattice in the form of a $1/L$ background density.

Define $P_B=|B\rangle\langle B|$, as the projection operator onto the
bound state, and rewrite
Eq.(\ref{correl}), as
\begin{equation}
S=\sum_{x=1}^D{\cal G}_{+-}(x)=\frac{z^2}{\lambda_B^2} \langle B |
G_+   \sum_{x=1}^D \Big[\Big(\frac{M}{\lambda_B}
\Big)^{x-1} \!\!\!\!\! - P_B   \Big] G_- | B \rangle.
\end{equation}
The bound state does not contribute to the correlators inside the
sum. Therefore
we can project out the bound state from $M$ and then perform the summation
\begin{eqnarray}
S
&=&\frac{z^2}{\lambda_B^2} \langle B | G_+ \Big[  \sum_{x=1}^D
\Big(\frac{M}{\lambda_B}
- P_B \Big)^{x-1}  - P_B   \Big] G_- | B \rangle\nonumber\\
&=&\frac{z^2}{\lambda_B^2}\langle B | G_+ \Big[
\frac{1}{1-\frac{M}{\lambda_B}+ P_B}  - P_B   \Big] G_- | B\rangle.
\end{eqnarray}
(The single $P_B$ outside the summation originates from the  $x=1$
contribution.)
We can remove  $G_+$ and $G_-$ from the above equation, because  the
bound state is also an eigenstate of the
lowering operator, $G_+|B\rangle=\lambda_p |B\rangle$,
\begin{equation}
S= \frac{|\langle B|0 \rangle|^2}{\lambda_B^2} \langle 0 |  \Big[
\frac{1}{1-\frac{M}{\lambda_B}+ P_B}  - P_B    \Big] | 0\rangle
\label{S-reduction}
\end{equation}
writing $zG_\pm= M -G_0- zG_\mp$, and using that
$G_0=|0\rangle\langle 0|$ is the projection operator onto the first site,
and also that $M-\lambda_B P_B$ has no projection onto $|B\rangle$.

The sum rule we seek is now reduced to the identity
\begin{equation}
\sum_{k \neq B} \frac{ |\langle 0|k\rangle|^2}{\lambda_B-\lambda_k}
=\frac{z^2\lambda_p^2}{\lambda_B \lambda_0}
\label{Sumrule}
\end{equation}
with $\lambda_0 =|\langle B|0\rangle|^2= \langle B|G_0|B\rangle$
$=\langle B|M -z(G_++G_-)|B\rangle$ $=\lambda_B-2z\lambda_p$.
The left hand side is easily evaluated, using
Eq.(\ref{h-zero}) and that $\lambda_B-\lambda_k =z^2-2z \cos k+1$,
\begin{equation}
\sum_{k \neq B} \frac{ |\langle 0| k \rangle|^2}{\lambda_B-\lambda_k}
= \frac{2}{\pi} \!\! \int_0^\pi \!\! dk \frac{z^2\sin^2 k}{(z^2-2 
z\cos k+1)^2}.
\end{equation}
This is an elementary contour integral along the unit circle in the
complex $w=e^{ik}$ plane,
with a double pole at $w=w_b=z$ within the circle in addition to a
single pole at $w=0$.
The integral is indeed equal to $z^2/(1-z^2)$, the right hand side of
Eq.(\ref{S-reduction})
($\lambda_p^2=\lambda_B=(1+z)^2$ and $\lambda_0=1-z^2$).

\subsection{Quadratic Algebra at  $r\neq p$} \label{algebra-r-neq-p}

The proof of perfect screening for general $r/p$ follows the same
pattern as at $r=p$.
The operators obeying the quadratic algebra conditions, Eq.(\ref{rqalgebra}),
are again expressed in terms of raising and lowering operators $L_\pm$,
\cite{Arndt98, Arndt99, Rajew00, Kafri02, Schutz03, Kafri03}
\begin{eqnarray}
G_+ &=& \frac{1}{a}[I+L_-+(a-1) G_0 + (s-1) G_0 L_-],\nonumber\\
G_- &= &\frac{1}{a}[I+L_++(a-1) G_0 + (s-1) L_+ G_0],\nonumber\\
G_0 &=& |0\rangle\langle0|,
\label{rnqp-operators}
\end{eqnarray}
where $a=r/p$ and $s^2=1-(a-1)^2$. The transfer matrix retains its form
\begin{eqnarray}
M = zG_++zG_-+G_0= \frac{4z}{a}\left( 1- \frac{1}{4}H\right)
\end{eqnarray}
with modified  Hamiltonian,
\begin{eqnarray}
H &\equiv& 2I -L_+ -L_- - \frac{1}{\tilde z}G_0 \nonumber\\
&&- (s-1)(G_0 L_- + L_+ G_0).
\end{eqnarray}
and with $1/\tilde z \equiv a/z +2(a-1)$.
This is again a one-dimensional single particle hopping process in a
half-space,
$n=0,1,2,\cdots$.  Compared to Eq.(\ref{hamil}) for $a=\frac{r}{p}=1$,
the  attractive potential at site $n=1$ deepens for $r>p$ (reducing
the clustering and
correlation lengths). The novel element is the modified hopping probability
$s$ between
sites $n=1$ and $n=0$.
There is still only one bound state
\begin{eqnarray}
\phi_n &=& \frac{1}{\sqrt{Z_B}} ~w_b^{n} \quad \mbox{for} \quad n
\geq 1,\nonumber \\
\phi_0 &=& \frac{1}{\sqrt{Z_B}}~ \frac{1}{s},\nonumber\\
\frac{1}{\tilde z}& =&\frac{1}{w_b}+ (1-s^2)w_b,
\label{pnqr-bound}
\end{eqnarray}
with the same functional form for the bound state energy as before,
Eq.(\ref{bound-energy}).

The derivation of the extended states is also straight forward.
They are again of the form, Eq(\ref{scatter}),
with the same energies, $E_k=2(1-\cos k)$, but
satisfying the modified relations
\begin{equation}
sA_0(k)= \cos(\theta_k) =\frac{s^2\cos(\theta_k+k)}{2\cos k-
\frac{1}{\tilde z}}.
\end{equation}
This leads after some algebra to
\begin{equation}
\phi_0^2=\frac{(w^\prime_bw_b-1)\sin^2 k~~2/D}{(w_b+w_b^{-1}-2\cos
k)(w^\prime_b+{w^\prime_b}^{-1}-2 \cos k)}.
\end{equation}
$w^\prime_b$ is the second root of the bound state equation
Eq.(\ref{pnqr-bound}).

\subsection{Quasi-Particle Representation}\label{quasi-r-neq-p}

We can now identify the exact form of the ratio $\alpha/\beta$ in the
quasi-particle representation, eq.(\ref{quasip}).
The representation mixes the $G_\pm$ operators in Eq.(\ref{rnqp-operators})  as
\begin{equation}
G_p = \alpha G_+ + \beta G_-,~~~G_m= \beta G_+ + \alpha G_-.
\end{equation}
The projection operator, $G_0 = |0\rangle \langle 0 |$,
and the transfer matrix $M$ are invariant.
The latter implies $\alpha + \beta=1$.

The quasi-particle two-point correlation functions take the same form
as the particle correlators at $r=p$.
In particular,  the quasi-particle correlation function  ${\cal
G}_{pm}(x)$ is zero for all $x<0$.
This is true when the bound state is also an eigenstate of $G_p$:
\begin{equation}
G_p | B \rangle = \lambda_p | B \rangle.
\label{one-side}
\end{equation}
Inserting the bound state, Eq.(\ref{pnqr-bound}), yields
\begin{equation}
a\lambda_p = 1+ \alpha w_b + \frac{\beta }{w_b}
\end{equation}
and
\begin{equation}
\frac{\beta}{\alpha} = \frac{(s^2-1)w_b-(1-a)}{1/w_b+(1-a)}.
\label{b-ov-a}
\end{equation}
The lines of constant $\beta/\alpha$ are shown in Fig.(\ref{mixing-strength}).
(Insert the above equations for $\omega_b$, $\tilde z$, and the
relation between $a$ and $s$.)
The contours  coincide with our numerical results.

\subsection{Perfect Screening at $r\neq p$} \label{perfect-screening-r-neq-p}

The final step is to prove perfect screening in terms of the quasi-particles:
\begin{equation}
S=\sum_{x=1}^\infty {\cal G}_{pm}(x) = -{\cal G}_{pm}(0).
\label{perfect-screen}
\end{equation}
The left hand side  reduces to exactly the same form as Eq.(\ref{S-reduction}),
using the exact same steps, because the bound state is  an eigenstate of $G_p$
just like the particle operator $G_+$ at $r=p$; that is all we used there.
The right hand side is different, because $\langle n_p
n_m\rangle=2\alpha\beta \rho$ is not zero anymore.
Since $\alpha+\beta=1$, it is still true that
$\rho_p=\rho_m=\rho_+=\rho_-=z\lambda_p/\lambda_B$.
Therefore the sum rule equation, Eq.(\ref{Sumrule}) now takes the form
\begin{equation}
\sum_{k \neq B} \frac{ |\langle 0|k\rangle|^2}{\lambda_B-\lambda_k}
=\frac{1}{\lambda_0}\left(\frac{z^2\lambda_p^2}{\lambda_B}
-2\alpha\beta z\lambda_p\right)
\label{rnpSumrule}
\end{equation}
with, as before, $\lambda_0 =|\langle B|0\rangle|^2= \langle B|G_0|B\rangle$
$=\langle B|M -z(G_p+G_m)|B\rangle$ $=\lambda_B-2z\lambda_p$.
The summation on the left leads again to a  $w=e^{ik}$ type
contour integral.
It still has only two poles inside the unit circle:
one double pole at $w=w_b$ and one single pole at $w=1/w^\prime_b$
(with $w^\prime_b$ the second root of Eq.(\ref{pnqr-bound}).)
The result is indeed equal to the right hand side after
inserting the proper expressions for the various eigenvalues and
some not very pretty algebra.

\bigskip
\section{Results and Conclusions}
\label{conclusion}

We have studied the two-species asymmetric exclusion process (ASEP)
to determine whether the addition of a local conservation law changes
the dynamic scaling properties.
In the Burgers (hydrodynamics) context the process conserves both momentum and
density. In the KPZ context it represents interface growth where the
numbers of up and down steps
are conserved. In the ASEP context  the particle numbers of both
species are conserved.

We find that the dynamic scaling exponent retains the KPZ $z=3/2$
value.   The AHR process factorizes at scales larger than the  clustering
length scale, $\xi$, into two-independent KPZ processes.
At $r=p$, where the passing and hopping probabilities are equal,
this factorization occurs in terms of $+$ and $-$ particles, while at
$r\neq p$ it is established in terms of quasi-particles.  This
factorization expresses itself  as
perfect screening between the two species of quasi-particles.  $\xi$,
the screening length, coincides with the clustering length scale and
represents the crossover length scale between single KPZ scaling 
(within each cluster) and factorized
$(\mbox{KPZ})^2$ type scaling.

The conventional method for measuring the dynamic exponents in
simulations in terms of the time
evolution of the interface width fails in this process due to the
presence of time-oscillations
with a period proportional to the system size; quasi-particles fluctuations
have  non-zero and opposite
drift velocities. Instead, we determined the dynamic scaling from the
two-point correlation functions.
This might be the first time that it is done in this manner.

The stationary state of this process has been studied extensively in
the recent literature, because it
obeys the so-called matrix product ansatz (MPA). We used this to
prove rigorously
the factorization of the fluctuations in terms of quasi-particles.
This previously unknown feature of the algebraic structure of the MPA method
needs to be understood better,
in particular in the context of clustering  phenomena in general.

The perfect screening phenomenon is clearly a topological feature.
The above presentation only partially  exposes those topological properties;
by bringing the perfect screening condition into the form of
Eqs.(\ref{Sumrule}) and (\ref{rnpSumrule}).
The right hand sides of both equations only involve bound state
properties. Their left hand sides however
involve a summation over all extended states; i.e., their
projections onto $n=0$ ($|\langle 0|k\rangle|^2$).
The poles of the contour integral links this to the bound states
and the quasi-particle mixing.
The formulation of a general proof is important  because, if
topological, the prefect screening and (KPZ)$^2$ scaling
at large length scales will be generic features, valid to many more
processes with clustering.
Its limitations can teach us when and how novel type  dynamic scaling sets in.

\emph{Acknowledgment --}
This research is supported by the National Science Foundation under
grant DMR-0341341.


\begin{references}
\bibitem{Leven} The Levenberg-Marquardt method is used with a 
confidence bound of 95\%.
\end{references}
\end{document}